# Hafnia for analog memristor: Influence of stoichiometry and crystalline structure

Li-Heng Li,[1,2] Kan-Hao Xue,[1,2*] Jun-Hui Yuan,[1,2] Ge-Qi Mao,[1,2] and Xiangshui Miao[1,2]

[1]School of Integrated Circuits, School of Optical and Electronic Information, Huazhong University of Science and Technology, Wuhan 430074, China

[2]Hubei Yangtze Memory Laboratories, Wuhan 430205, China

***Corresponding Author**, E-mail: xkh@hust.edu.cn (K.-H. Xue)

**Abstract**

The highly non-linear switching behavior of hafnia memristor actually hinders its wide application in neuromorphic computing. Theoretical understanding into its switching mechanism has been focused on the processes of conductive filament generation and rupture, but possible phase transition and crystallization around the region of conductive filaments (CFs) due to the variation of O content have been paid less attention to. In this paper, $HfO_x$ structural models covering the full stoichiometries from Hf to $HfO_2$ were established, and the crystal structure evolution during the reduction process of hafnia was obtained through first-principles calculation. The electronic structures and O vacancy migration characteristics of these structures were analyzed. A criterion was prescribed to predict the mode of abrupt binary switching or gradual conductance modulation according to the structure evolution of the CFs. In particular, factors that influence the merging of tiny conductive channels into strong filaments are intensively discussed, including the anisotropy of O vacancy migration and the size effect. The feasibility of Mg doping to achieve robust gradual switching is discussed.

# 1. Introduction

Due to their high speed, CMOS-compatible nature and 3D integration potential [1–4], binary oxide memristors have been widely used in non-volatile data storage, logic computing and neuromorphic computing [5–12]. In particular, hafnia ($HfO_2$)-based memristors attract great attention for their downscale potential, nanosecond-level erasing time and excellent data retention [13–16]. Moreover, the resistive switching (RS) window of hafnia-memristors is typically higher than other binary oxides such as $TaO_x$ and $TiO_x$ [17–21], favoring their application as binary data storage. The capability of establishing a robust Schottky barrier between the conductive filaments (CFs) and $HfO_2$ has been revealed to account for such large RS window [22]. However, the actual application scenario may require different RS characteristics. For the sake of neuromorphic computing, linearity and symmetry are key aspects of the analog RS of the memristor [23]. In general, its highly nonlinear RS behavior hinders the application of hafnia-based memristor in neuromorphic computing [24,25]. Element doping and combined use with other RS materials are possible solutions, but there are still no well-established theory to overcome the abrupt RS behaviors in $HfO_2$.

To fundamentally resolve this problem, it is necessary to gain a deeper and material-specific understanding into the RS mechanism of $HfO_2$-based memristor. While the commonly accepted RS picture in $HfO_2$-based memristors is the formation and rupture of O-deficient CFs [26,27], exciting new discoveries have been made in recent few years. Celano *et al.* directly observed the conical structure of CFs in an $Hf/HfO_2/TiN$ memristor through their newly designed scalpel-SPM technique, which is a kind of conductive atomic force microscopy, and defined the CF as a local region populated with O vacancies ($V_O$'s) [28]. The cross sectional area of a typical CF in their experiment was ~40 $nm^2$ at the active Hf electrode size but merely ~10 $nm^2$ at the more inert TiN electrode side. Therefore, $V_O$'s diffuse and migrate from the active electrode side toward the inert electrode side during the formation of CFs. Although the exact composition/stoichiometry of the CF was not obtained, this work has lent strong support to the $V_O$-induced CF conduction in hafnia memristors. Kumar *et al.* observed that the CF in hafnia consists of an O-poor conductive core surrounded by an O-rich region [29]. The formation and dissolution of the CFs were ascribed to the radial thermophoresis and Fickian diffusion of O atoms driven by Joule heating. The role of O migration and its induced electrochemical reactions were further probed directly by Yang *et al.* in $HfO_2$ and $TaO_x$ memristors with spherical aberration (Cs)-corrected transmission electron microscope (TEM) [30]. In 2018, Yin *et al.* [31] showed high resolution TEM (HRTEM) evidences that the CFs in hafnia memristors consist of crystalline regions of monoclinic and orthorhombic symmetry, even though the pristine hafnia film was amorphous before electroforming. The work pointed out the significance of crystalline $HfO_x$ phases in the search of CF structure in amorphous hafnia-based memristors. For polycrystalline hafnia films, it has been shown that the CFs prefer to emerge near grain boundaries [32] where creation of $V_O$'s can be easier than in the bulk [33].

Very recently, Zhang *et al.* [34] identified the crystalline structure of the CFs in hafnia memristors through comprehensive HRTEM imaging for $Pt/HfO_2/Pt$, $TiN/HfO_2/Pt$, $Ta/HfO_2/Pt$, $Hf/HfO_2/Pt$ and $Ti/HfO_2/Pt$ cells. The CF consists of a hexagonal $Hf_6O$ metal core surrounded by an O-rich monoclinic $HfO_x$ (*m*-$HfO_x$) or tetragonal $HfO_x$ (*t*-$HfO_x$) shell. The crystalline feature of the CFs is very prominent, though the pristine hafnia thin films were always amorphous. In that work, the electroforming and SET/RESET operations were thorough as to support a $10^6$ RS window, thus the core-shell CF structure has exactly been confirmed at least for the binary RS mode of hafnia memristors. The observation of the thermodynamically stable $Hf_6O$ phase (*i.e.*, ~$HfO_{0.17}$) is also consistent with the theoretical prediction by McKenna ($HfO_{0.2}$) [35] through an energetic argument.

Notwithstanding these new discoveries including concrete TEM evidences, it needs to be emphasized that to identify the CF of hafnia memristors under gradual switching mode (or say, working as an analog memristor) through TEM is much more difficult than under the binary switching mode. A gradual resistance modulation may not alter the symmetry of the CF, but only involves mild ionic migration. The conductivity variation among different regions is also expected to be much slighter than the binary mode. Since the gradual RS mode is very important for neuromorphic computing applications, theoretical calculations become a viable choice. In particular, previous discovery that the CF involves crystalline regions could greatly facilitate the theoretical investigation on the CF composition and structure in hafnia memristors under the gradual RS mode. Using first-principles calculations, the RS mechanism in hafnia has been intensively studied. Perevalov *et al.* used first-principles calculations to simulate the electronic structures of various defects in HfO$_2$, including the V$_O$, interstitial O, interstitial Hf, as well as Hf substituting for O, and showed that V$_O$'s are the key defects for the charge transport and RRAM operability [36], verifying the importance of Vo in the RS process. The theoretical study on the formation and migration of Vo shows that there is an optimal path with low migration barrier [37], which is beneficial to the reduction of operating voltage and the acceleration of the signal response. However, low migration barrier also leads to poor thermal stability of HfO$_2$-based memristors [38]. Several works have shown the tendency of O vacancy clustering in HfO$_2$ [39,40], which explains the formation of conical structure CFs in HfO$_2$ and confirms a size limit of CFs in Hf or Hf-like compositions (~0.4 nm$^2$) [40].

These previous theoretical researches, however, did not cover all possible crystalline HfO$_x$ phases between Hf and HfO$_2$. Most works emphasize the significance of V$_O$ concentration for conduction, but the resulting phases with sufficient V$_O$ concentration that act as CFs are unclear. On account of the experimental identification of crystalline CF structures, various HfO$_x$ (0≤*x*<2) phases ought to be examined for possible CF composition of hafnia memristors under the gradual RS mode. Actually, a number of Hf sub-oxide phases have been either predicted or discovered. Some well-known hexagonal (*h*-) phases can be regarded as the derivative of *h.c.p.* Hf, including *h*-Hf$_6$O ($R\bar{3}$), *h*-Hf$_3$O ($R\bar{3}c$) and *h*-Hf$_2$O ($P\bar{3}1m$) [41]. The ground state of HfO has been predicted to be tetragonal (*t*-) $I4_1/amd$, which is metallic and has also been extensively observed in the experiments by Zhang *et al.* [34]. Ground state ZrO, however, was predicted to possess a $P\bar{6}2m$ structure [42]. Its *h*-HfO counterpart is energetically less favorable than *t*-HfO, and surprisingly shows a tiny band gap [43]. Predicted Hf$_2$O$_3$ phases involve a tetragonal $P\bar{4}m2$ structure [44] and an orthorhombic *Ibam* structure [45], with the former slightly lower in energy. Several other pressure-induced HfO$_x$ sub-oxides have also been predicted [41]. The sequence of sub-oxide presence during the reduction of HfO$_2$ has been summarized as roughly from the tetragonal symmetry to the hexagonal symmetry [46]. A very recent work shows that the transitional stoichiometry is at around HfO$_{0.7}$ [47].

In hafnia-memristors, the reduction of HfO$_2$ is limited to directional electric field and thermal forces, which is different from the reduction in bulk forms. We shall establish the relation among various HfO$_x$ phases in memristor applications. Based on the comprehensive materials information, we then derive the working modes of hafnia memristors. The size-composition relation of CFs in hafnia will also be investigated.

## 2. Modeling and Computational Details

### 2.1 General considerations

At atmospheric pressure and low temperature, monoclinic $P2_1/c$ is the most stable phase of $HfO_2$, which can stand to up to ~2000 K [48], thus becoming the natural choice of our starting point for the pristine state of $HfO_2$-based devices. As only the CFs and their surrounding regions are significant, which account for the RS behaviors, such crystalline starting point also makes sense for the initially amorphous $HfO_2$ memristors. There is concrete experimental evidence that the CF-surroundings can be crystallized through the Joule heating effect from electroforming, SET and RESET operations [34]. Therefore, we gradually increased the oxygen vacancy concentration in $m$-$HfO_2$, covering the full stoichiometries from $HfO_2$ to Hf. After structural optimization, these reduced $HfO_x$ models could give useful information regarding the relatively stable $HfO_x$ structures for $0<x<2$, which can possibly be meta-stable from thermodynamics but are more favorable against other structures at that specific stoichiometry. Moreover, ion migration barriers were systematically calculated to explore the possible transition routes between these stable or meta-stable phases. The kinetic information is highly relevant because our primary concern is the scenario of memristor application, where only uni-directional (though can be positive or negative) electric field is applied to the capacitor. In this case some thermodynamically favorable phases may not be reached due to high transitional barriers.

### 2.2 Modeling details

The structure of $m$-$HfO_2$ is illustrated in **Figure 1**, which derives from the cubic fluorite structure, but the Hf coordination has been reduced from 8 to 7 in terms of the monoclinic distortion. There are two non-equivalent O sites, 3-coordination O(A1-A4) and 4-coordination O(B1-B4). The O-atom layer including O(B1), O(B2), O(A3) and O(A4) is equivalent to the layer including O(A1), O(A2), O(B3) and O(B4). A specific amount of O atoms (from 1 to 8) was removed from the 12-atom primitive cell $Hf_4O_8$, and the remaining structure was subject to full relaxation. Of course, we cannot start from all possible $m$-$HfO_2$ supercells for O removal, which is an endless task. Nevertheless, previous experience shows that the $Hf_4O_8$ primitive cell is sufficiently large to yield possible sub-oxide phases. Moreover, we shall show that all reported or predicted $HfO_x$ cases are included in or related to our search results, which proves the effectiveness of the method *a posteriori*. Careful symmetry analysis has eliminated a number of redundant cases, ending up with only 32 inequivalent $HfO_x$ models, as summarized in **Figure 2**.

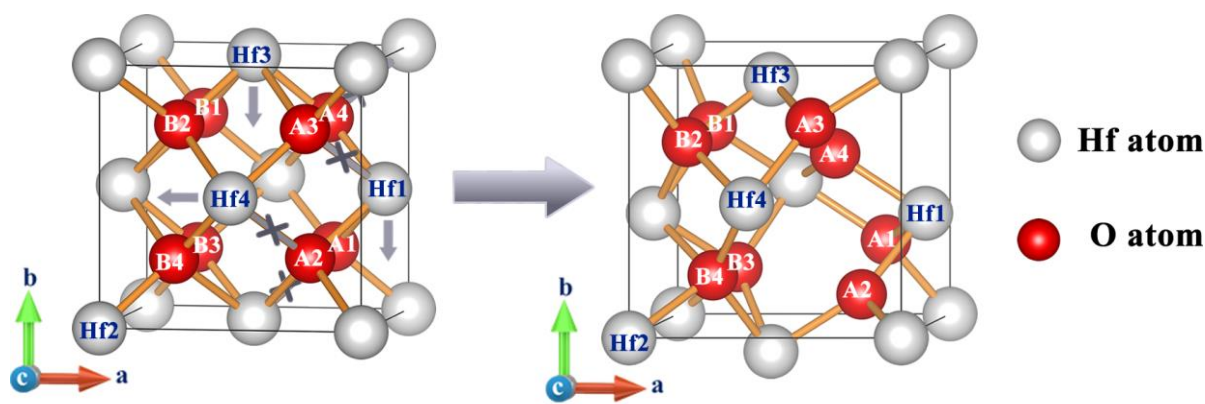

**Figure 1**. The structure of $m$-$HfO_2$ (right) as derived from the cubic fluorite structure (left). The basic directions of Hf movement and the broken Hf-O binds are marked for thr monoclinic distortion.

**2.3 Computational details**

The *ab initio* calculations were carried out using density functional theory (DFT) [49,50] as implemented in the plane-wave based Vienna *Ab initio* Simulation Package (VASP 5.4.4) [51,52]. It is well known that DFT-GGA systematically underestimates the band gaps for semiconductors and insulators. Hence, we shall implement an efficient self-energy correction scheme (DFT-1/2) [53], originally proposed by Ferreira *et al.*, and later improved by Xue *et al.* (shDFT-1/2) [54], to carry out electronic structure calculation. Both the electronic structure and the Vo migration barrier were calculated based on 2×2×2 supercells. The migration barrier was calculated using a climbing image nudged elastic band (CI-NEB) method, developed by Henkelman and Jónsson [55].

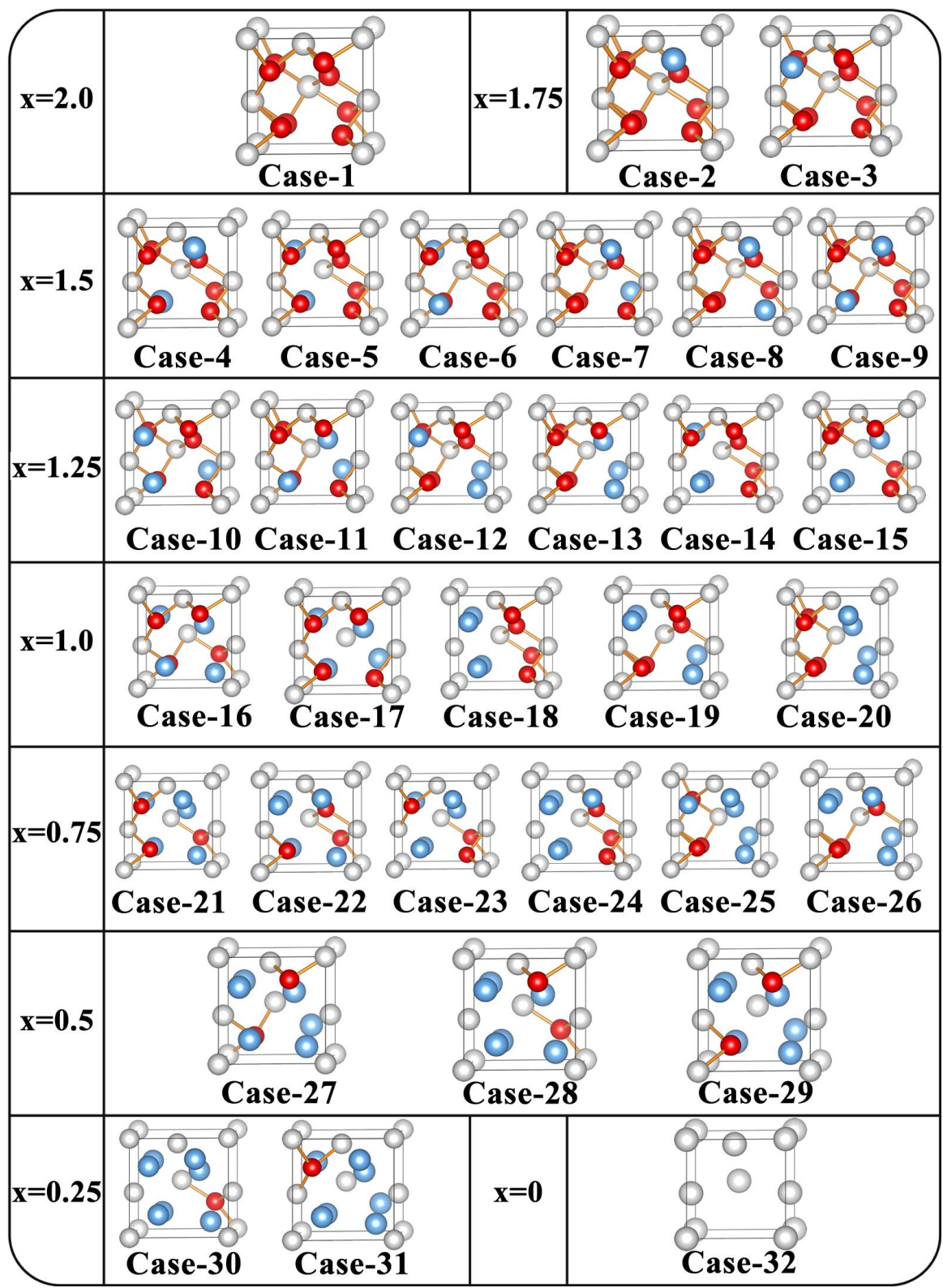

**Figure 2**. The 32 inequivalent unit cell models of the full range of Hf-O binary compounds.

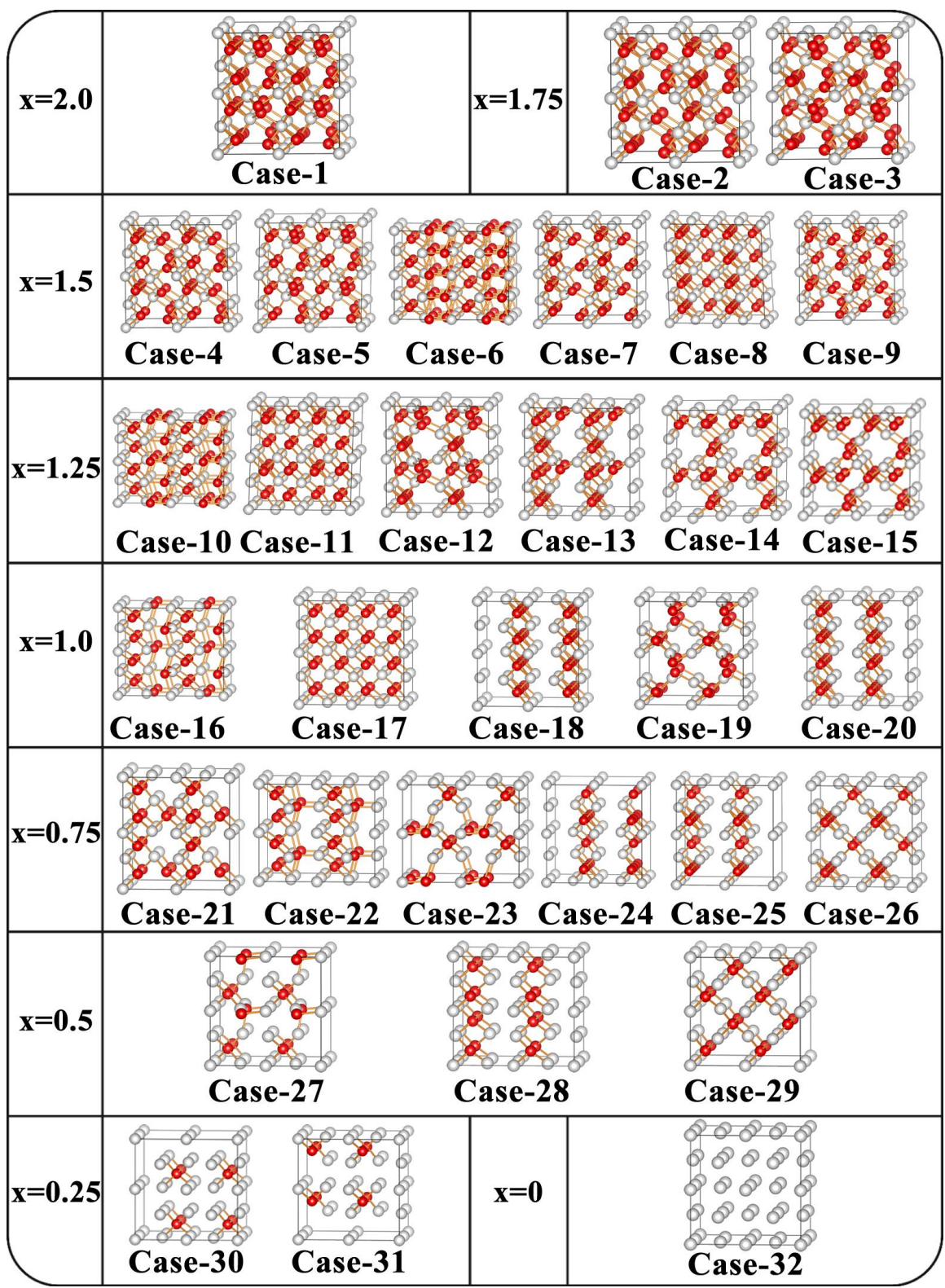

**Figure 3**. Structural demonstration of the 32 inequivalent models after complete relaxation.

## 3. Results and Discussion

### 3.1 Structural analysis of HfO$_x$ at various stoichiometries

For structural optimization of the 32 primitive cell models, high precision full relaxation was carried out first. In order to visualize the changes of crystal structures, the relaxation results were extended to 2×2×2 supercell structures, as illustrated in **Figure 3**. On the other hand, **Figure 4** summarizes the possible HfO$_x$ structures under various stoichiometries, which may be classified into three regimes.

- Regime 1 (1.5<$x$≤2.0):

  All HfO$_x$ structures in this region remain in the monoclinic symmetry ($m$-HfO$_x$), including Case-2 and Case-3.

- Regime 2 (1≤$x$≤1.5):

  There are both monoclinic phases ($m$-HfO$_x$: Case-4, Case-5, Case-7 and Case-9) and tetragonal phases ($t$-HfO$_x$: Case-6 and Case-8). The binding energies of $m$-HfO$_x$ structures are between -7.58 eV/f.u. and -7.70 eV/f.u. (see **Supplementary Note 1**), where the lowest-energy structure is Case-4, which will subsequently be intensively analyzed as the prototype $m$-HfO$_{1.5}$ structure. The $t$-HfO$_x$ structure is shown in **Figure 4(b)**, where Hf atoms maintain $f.c.c.$-like structure (though distorted) and O atoms can occupy two types of positions (Site-1 and Site-2). When all O atoms occupy Site-1, the structure is the $P4_2/nmc$-HfO$_{1.5}$ phase (Case-8). Provided that O atoms occupy both Site-1 and Site-2, a $P\bar{4}m2$-HfO$_{1.5}$ phase (Case-6) can then be recovered. The obtained HfO$_{1.25}$ structures are similar to that of HfO$_{1.5}$, and they can also be divided into $m$-HfO$_x$ (Case-12) and $t$-HfO$_x$ (Case-10 and Case-11).

- Regime 3 ($x$<1.0):

  There are a variety of phase structures in this regime, but in all cases the Hf atoms remains in quasi-$f.c.c.$ structure and the O atoms occupy Site-1 or Site-2 positions. Therefore, these structures can be regarded as O-deficient structures of $P4_2/nmc$-HfO$_{1.5}$ and $P\bar{4}m2$-HfO$_{1.5}$.

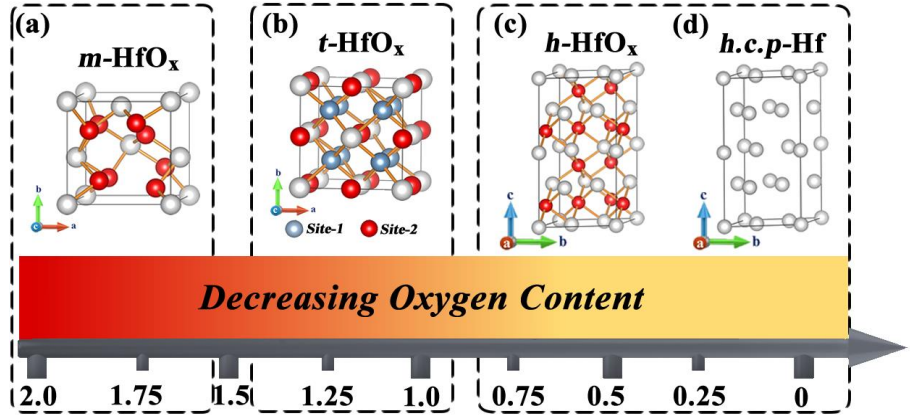

**Figure 4**. The phase structure evolution of $HfO_x$ during the oxygen content decreasing.

However, no known $h$-$HfO_x$ structure was obtained through this modeling method. This is due to the constriction of the Hf-subsystem topology, which fails to be converted to hexagonal symmetry. Previous studies [41,45] have demonstrated the vicinity of pure Hf in the Hf-O system, where the $h.c.p.$ structure is preserved and the extra content of oxygen is distributed among the interstitial positions octahedrally coordinated by Hf atoms. It has also been confirmed experimentally that $HfO_x$ will be converted into $h$-$HfO_x$ at x=0.7 [47]. Theoretically, the O content of $h$-$HfO_x$ can reach $x$=0.83, and the corresponding structure is $h$-$Hf_6O_5$, as shown in **Figure 4(c)**. Therefore, the $HfO_x$ structures in Regime 3 ($x$<1.0) is considered to be $h$-$HfO_x$. Our method thus covers the O-rich phases prior to the tetragonal-to-hexagonal transition.

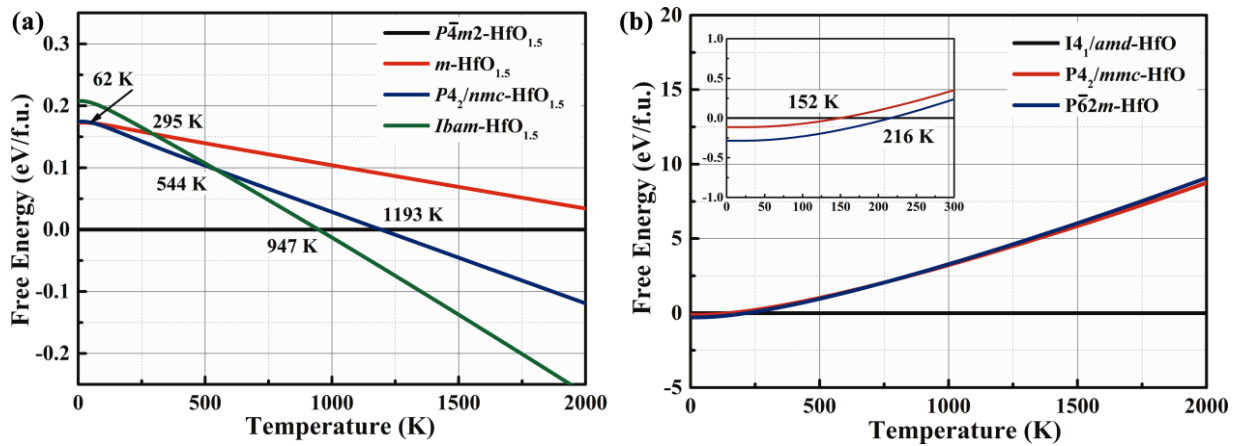

**Figure 5**. The Helmholtz free energies of (a) $m$-$HfO_{1.5}$, $P4_2/nmc$-$HfO_{1.5}$ and $P\bar{4}m2$-$Hf_2O_3$ and (b) $I4_1/amd$-HfO, $P4_2/mmc$ –HfO, and $P\bar{6}2m$-HfO.

Obviously, $HfO_2$ underwent a transition from the monoclinic phase to tetragonal phases,

and eventually to hexagonal symmetry upon decreasing the O content. A critical stoichiometry is discovered to be $x=1.5$, at which there are three phases, $m$-HfO$_{1.5}$, $P\bar{4}m2$-HfO$_{1.5}$ (hereafter named α-HfO$_{1.5}$ or α-Hf$_2$O$_3$), and another tetragonal phase of HfO$_{1.5}$ derived from $P4_2/nmc$-HfO$_2$. Their crystal structures are illustrated in **Figure 3** as Case-1, Case-6 and Case-8. Another predicted *Ibam* phase of HfO$_{1.5}$ (referred to as β-HfO$_{1.5}$ or β-Hf$_2$O$_3$ for convenience) [45] is not included in **Figure 3**, but its relation to other HfO$_x$ phases will be clear soon. The tetragonal phase of HfO$_{1.5}$ derived from $P4_2/nmc$-HfO$_2$ has a $P2/c$ symmetry, and will be referred to as γ-HfO$_{1.5}$ in this work. **Figure 5(a)** shows their Helmholtz free energy variations with respect to temperature $T$, with the vibrational entropies calculated using the harmonic oscillator model (details given in **Supplementary Note 2**). Zero-point energy was added to the free energy for $T = 0$ K in each phase. With the lowest free energy at zero temperature, α-HfO$_{1.5}$ is used as a reference. The results show that the phase transition temperature between $m$-HfO$_{1.5}$ and α-HfO$_{1.5}$ is 2490 K, and that between α-HfO$_{1.5}$ and γ-HfO$_{1.5}$ is 1193 K. These high temperatures can hardly be obtained during the memristor operations. In other words, reversible phase transition between α-HfO$_{1.5}$ and the other two phases is unlikely through Joule heating. Nevertheless, the phase transition between γ-HfO$_{1.5}$ and $m$-HfO$_{1.5}$ can occur at merely 62 K. This is a remarkable result, since the phase transition temperature between $t$-HfO$_2$ and $m$-HfO$_2$ is well known to be as high as 1973 K [48]. Hence, the existence of abundant O vacancies improves the stability of $P4_2/nmc$-like HfO$_x$ [34]. It also implies that a CF with $m$-HfO$_{1.5}$ structure would spontaneously transform into the γ-HfO$_{1.5}$ structure.

A different Hf$_2$O$_3$ phase with the *Ibam* symmetry (*i.e.*, β-HfO$_{1.5}$ in this work) was predicted by Rushchanskii *et al.* in 2018 [45]. **Figures 6(a) and 6(b)** compare the atomic structures of β-HfO$_{1.5}$ and γ-HfO$_{1.5}$. Their topological structures are extremely similar. Actually, β-HfO$_{1.5}$ is obtained by introducing $b$-direction V$_O$ chains in cubic HfO$_2$ ($c$-HfO$_2$), while γ-HfO$_{1.5}$ is derived through the same procedure, but starting from $t$-HfO$_2$. Therefore, O anions are strictly aligned along the $b$-direction in β-HfO$_{1.5}$, but a zigzag sequence of O anions is observed in γ-HfO$_{1.5}$. Our calculation roughly shows a critical temperature of 544 K for their phase transition. Hence, β-HfO$_{1.5}$ may be obtained from γ-HfO$_{1.5}$ at a finite temperature.

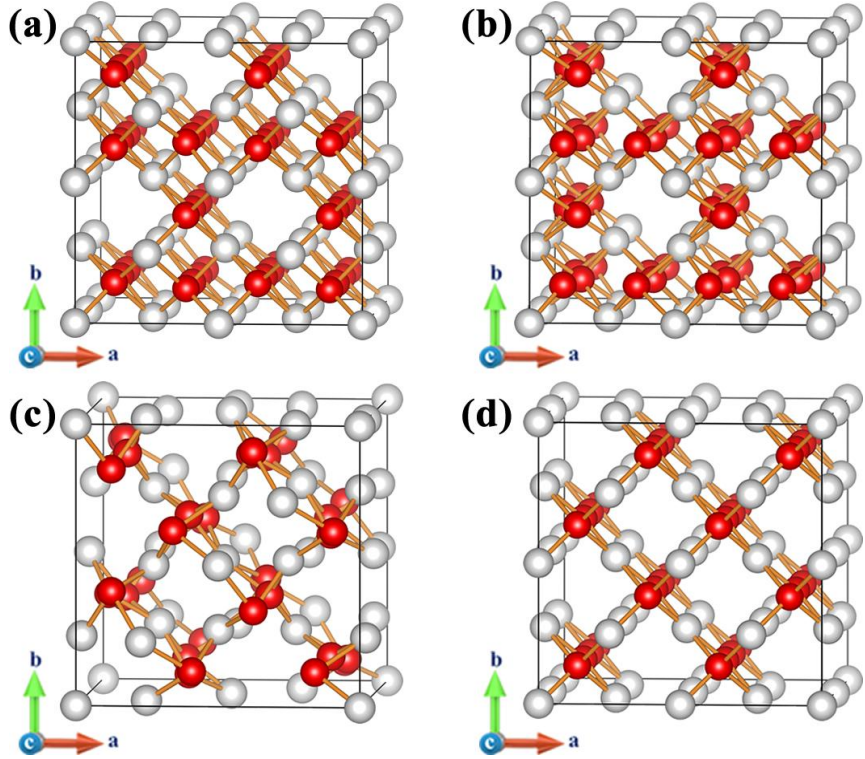

**Figure 6**. The structure diagrams of (a) *Ibam*-HfO$_{1.5}$ (β-HfO$_{1.5}$), (b) *P4$_2$/mmc*-HfO$_{1.5}$ (γ-HfO$_{1.5}$), (c) *I4$_1$/amd*-HfO (α-HfO) and (d) *P4$_2$/mmc*-HfO (γ-HfO).

Besides $x=1.5$, another important stoichiometry during the reduction process of hafnia is $x = 1$, where the lowest energy HfO phase is *I4$_1$/amd*-HfO as predicted by Zhu *et al*. [56] (structure shown in **Figure 6(c)**, with unit cell rotated by 45° along the *c*-axis to be better compared with the other HfO phase). It will be called α-HfO for convenience, while β-HfO refers to the $P\bar{6}2m$ phase [42,43]. Moreover, HfO may also be obtained through introducing more O vacancies from γ-HfO$_{1.5}$, yielding the tetragonal γ-HfO phase in a *P4$_2$/mmc* space group, as illustrated in **Figures 6(b)** and **6(d)**. The free energy comparison for these HfO phases are given in **Figure 5(b)**. Although the ground state total energy calculation favors α-HfO, after considering the zero-point energy, β-HfO becomes the most favorable phase at zero temperature. The phase transition temperature between α-HfO and β-HfO (γ-HfO) is 216 K (152 K). Hence, β-HfO and γ-HfO may be converted to α-HfO at room temperature.

**3.2 Relations between HfO$_2$ and several Hf sub-oxides**

The Hf sub-oxides with tetragonal or orthorhombic symmetries are all meta-stable at room temperature and atmospheric pressure, according to thermodynamic data. However, they may be stabilized in a memristor device due to mechanical confinement and/or under a certain stress. In memristors, the special situation is that these phases should be derived during the reduction process of $HfO_2$, mainly within the electroforming step. Therefore, before exploring the possibility of a sub-oxide to serve as the CF, it is a pre-requisite to understand its relation to $m$-$HfO_2$, or to fluorite $c$-$HfO_2$ because $m$-$HfO_2$ can be regarded as a distorted version of the fluorite structure.

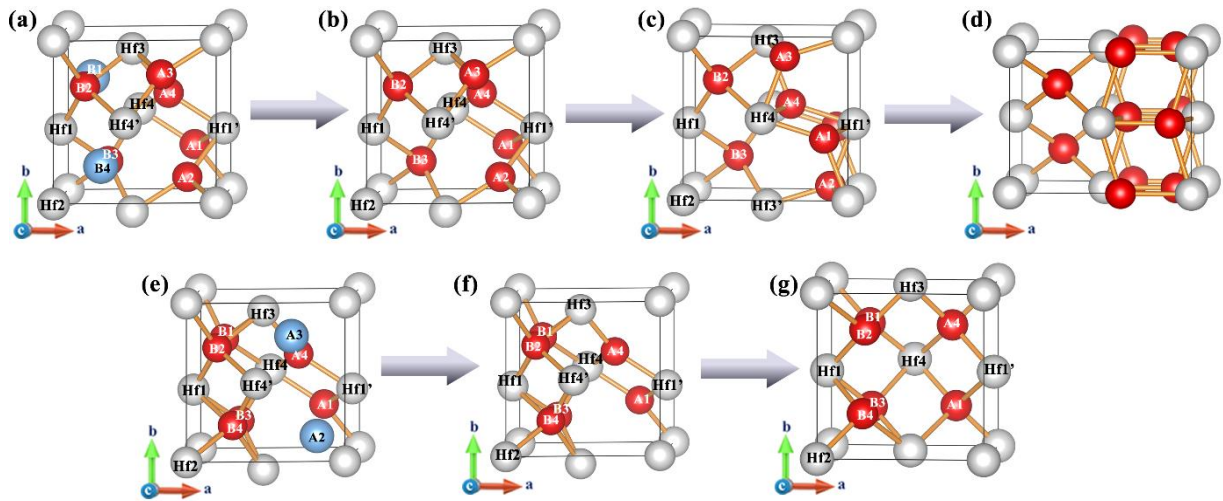

**Figure 7**. The process of atomic structure change between (a)-(d) $m$-$HfO_{1.5}$ and $P\bar{4}m2$-$HfO_{1.5}$, and (e)-(g) $m$-$HfO_{1.5}$ and $P4_2/nmc$-$HfO_{1.5}$.

To this end, we first note that it is the topology of the Hf subsystem that determines whether $HfO_x$ is in a tetragonal-like structure or a hexagonal structure. The Hf-subsystem is *f.c.c.* or distorted *f.c.c.* for the former, but becomes *h.c.p.* for the latter. The O anions may reside near the 8 fluorite sites (Site-1) or near the 4 rock salt sites (Site-2). Although $m$-$HfO_2$ is highly distorted, all its O anions still reside at Site-1 like in the fluorite structure. The monoclinic distortion reduced the Hf coordination number to 7, which better fits the ionic radius ratio between $Hf^{4+}$ and $O^{2-}$ [57]. As $x$ becomes 1.5, neither the fluorite structure nor the rock salt structure fits its stoichiometry, but their 1:1 mixture does. Hence, α-$HfO_{1.5}$ is in fact a natural candidate, which can be viewed as removing half of the 4-coordination O anions from $m$-$HfO_2$, followed by transferring the 3-coordination O anions from Site-1 to Site-2, as illustrated in **Figures 7(a)-7(d)**. The Site-2 O anions show a further distortion along $c$-axis, which renders 7-coordination for the Hf cations instead of 8. On the other hand, both β-HfO$_{1.5}$ and γ-HfO$_{1.5}$ only requires creating O vacancy chains along the *b*-axis of *m*-HfO$_2$, without interchanging Site-1 and Site-2 O anions, as illustrated in **Figures 7(e)-7(g)**. Actually, they possess no Site-2 O anions like in *m*-HfO$_2$ and *t*-HfO$_2$.

When it comes to HfO, we first notice that γ-HfO is closely related to γ-HfO$_{1.5}$, through simply introducing more O vacancies. At *x*=1.0, γ-HfO becomes much more symmetric (*P*4$_2$/*mmc*) than γ-HfO$_{1.5}$ (*P*2/*c*). On the other hand, α-HfO could also be derived from γ-HfO$_{1.5}$, with more O vacancy chains introduced. **Figures 6(c)** and **6(d)** show that the structures of α-HfO and γ-HfO are highly similar. The Hf and O atoms in α-HfO has slightly distorted compared with γ-HfO, but the global Hf-sub-lattice is still of *f.c.c.*-type and all O atoms reside at Site-1. Such tiny distortion reduces the free energy, rendering α-HfO more stable than γ-HfO at room temperature. Note that in all predicted and experimentally observed HfO phases, the O atoms occupy Site-1 rather than Site-2. As α-HfO$_{1.5}$ requires a partial occupation of Site-2, the rule observed in HfO confirms that α-HfO$_{1.5}$ is a bit isolated from the global path of hafnia reduction under the memristor situation, though the direct growth of HfO$_x$ under O-poor condition may yield α-HfO$_{1.5}$ [58,59].

If the *f.c.c.*-sub-lattice for Hf is to be maintained, the prototype models with O occupying Site-1 and Site-2 are *c*-HfO$_2$ and rock salt HfO, respectively. Their mixture generates HfO$_x$ with 1<*x*<2. The rock salt structure is indeed a simple potential phase for HfO where the O atoms may occupy and only occupy Site-2. However, besides it high energy, the phonon spectra of rock salt HfO show that it is kinetically unstable. The Hf-O bond length in rock salt HfO is also longer than in α-, β- and γ-phases of HfO, but it is more ionic than other phases. Detailed comparisons and calculation data for rock salt HfO are presented in **Supplementary Note 3**.

The next issue is the structure of HfO$_x$ with *x* < 1. The kinetic instability of rock salt HfO hinders the formation of cubic-like HfO$_x$ (*x*<1) by introducing O vacancies from rock salt HfO. However, introducing O vacancies in γ-HfO or α-HfO (which is almost the same) is a possible route. We therefore obtained γ-HfO$_{0.75}$, but its phonon instability (see **Supplementary Note 4**) indicates that keeping the *f.c.c.*-like Hf-sub-lattice is no longer feasible for highly reduced Hf sub-oxides with *x* < 1. Hence, a major structural change from *f.c.c.*-Hf to *h.c.p.*-Hf inevitably occurs near the stoichiometry of HfO. The experimentally observed transition stoichiometry at ~$HfO_{0.7}$ [47] is not far away. If starting from the *h.c.p.*-Hf sub-lattice side, the emergence of hexagonal β-HfO through backward oxidization still cannot be ruled out. To sum up, the relation between various Hf-O compounds during the reduction process of $HfO_2$ is given in **Figure 8**.

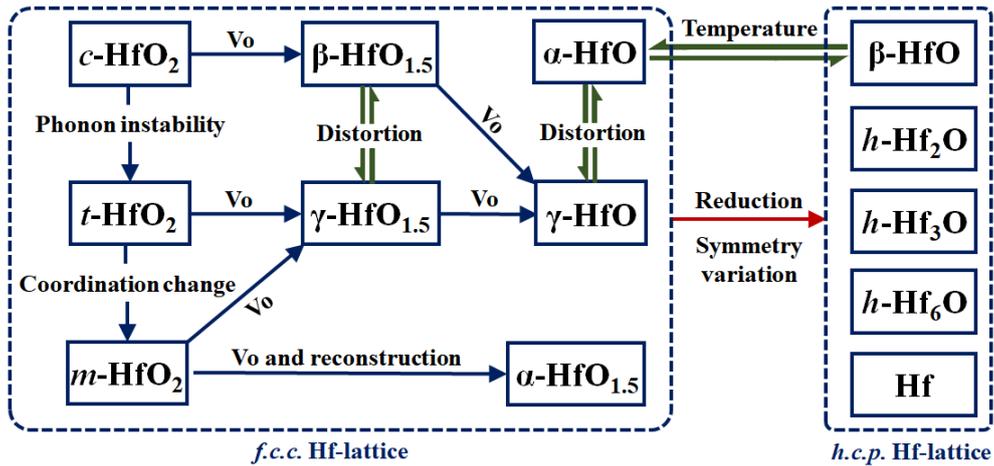

**Figure 8**. Relation of inter-transition between various Hf-O compounds during the reduction process of $HfO_2$.

### 3.3 The electrical and Vo migration properties in different $HfO_x$ phases

There are two reasons to understand the modes and difficulty of $V_O$ migration in these $HfO_x$ phases. On the one hand, it helps to determine whether a specific H-O sub-oxide phase may be obtained starting from $HfO_2$ (either amorphous or poly-crystalline), under the memristor operation conditions. This adds to our understanding of the CF formation from a kinetic point of view, besides the thermodynamic point of view. One the other hand, technically favorable CFs should possess proper migration barriers for Vo's, since too low barriers could deteriorate the stability of CFs and the device retention characteristics, but too high barriers would increase the device operation voltage.

#### 3.3.1 Transition barriers between various $HfO_x$ phases

To understand which sub-oxide phases are likely to emerge during the reduction process of hafnia, we set up a 2×2×2 supercell for *m*-HfO$_2$ to investigate the migration barriers of O vacancies, along B4→B1 (for the sake of α-HfO$_{1.5}$ formation) as well as A2→A3 (for the sake of γ-HfO$_{1.5}$ formation) paths, considering both neutral V$_O$ and V$_O^{2+}$. These paths are self-continuous as to enable drastic phase transitions. As illustrated in **Figure 9**, the migration barriers for V$_O$ and V$_O^{2+}$ are similar along the B4→B1 path, as high as ~5.0 eV. Yet, for the A2→A3 path, migration barriers for V$_O^{2+}$ and neutral Vo are merely 2.39 eV and 2.98 eV, respectively. Hence, under the driving force of electric field, it is easier to obtain γ-HfO$_{1.5}$ rather than α-HfO$_{1.5}$ from a kinetic point of view. On the other hand, to derive α-HfO$_{1.5}$ from γ-HfO$_{1.5}$, a part of O atoms have to migrate from Site-1 to Site-2, whose transition barrier is larger than migration from Site-1 to Site-1 (details shown in **Sect. 3.3.2**). Accordingly, the HfO$_x$ CF with 1.0<$x$<1.5 tends to keep the tetragonal symmetry as in γ-HfO$_x$.

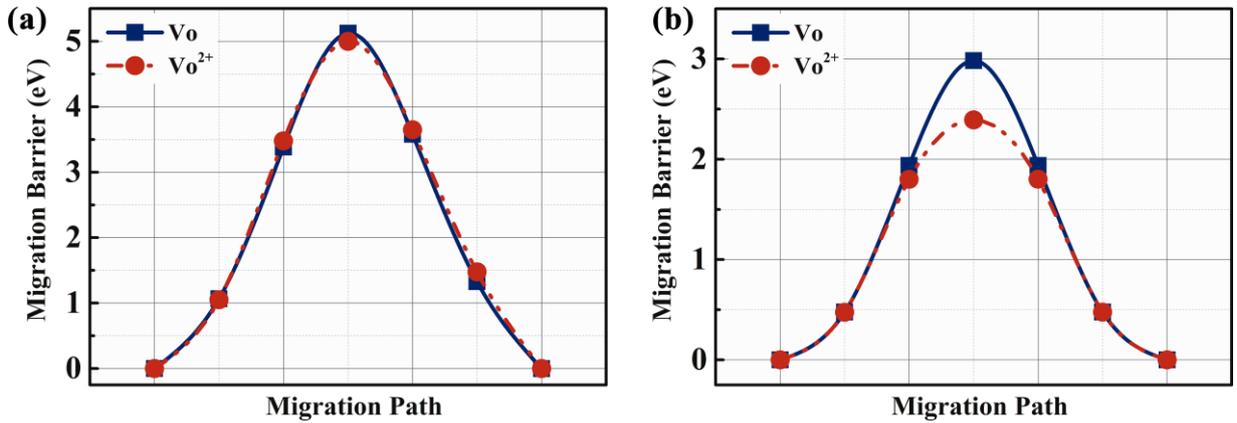

**Figure 9**. The Vo and Vo$^{2+}$ migration barriers in *m*-HfO$_2$ under the paths of (a) B4→B1 and (b) A2→A3.

### 3.3.2 RS within m-HfO$_x$

As mentioned in **Sect. 3.1**, when the Hf/O ratio $x$ is greater than 1.5, the single-phase HfO$_x$ crystal structure remains monoclinic. For such range of stoichiometry, the structures and density of states of *m*-HfO$_x$ are shown in **Figure 10**, confirming their semiconductor characteristic for the entire range 1.5<$x$<2.0. Provided that *m*-HfO$_x$ is to serve as the conductive channel, the RS will be confined to the high resistance state (HRS), but resistance modulation within a limited range (compared with the binary RS mode) is still possible. A potential benefit of this mode lies in that no sharp transition to metallic state is involved, therefore the linearity of the RS may become better. Indeed, as the O content decreases, the defect states in *m*-HfO$_x$ get stronger and accumulate below the CBM, lowering the energy gap to a value of 0.24 eV at *x*=1.5. The reduced defect excitation barrier should render much more excited carriers and improve the conductivity of *m*-HfO$_x$. The HRS of hafnia-based memristors may vary dramatically with O content. Our previous experiments demonstrate that this range is between 100 kΩ and 3 MΩ for a hafnia-memristor with 20 nm dielectric thickness and 80 μm × 80 μm device area [60].

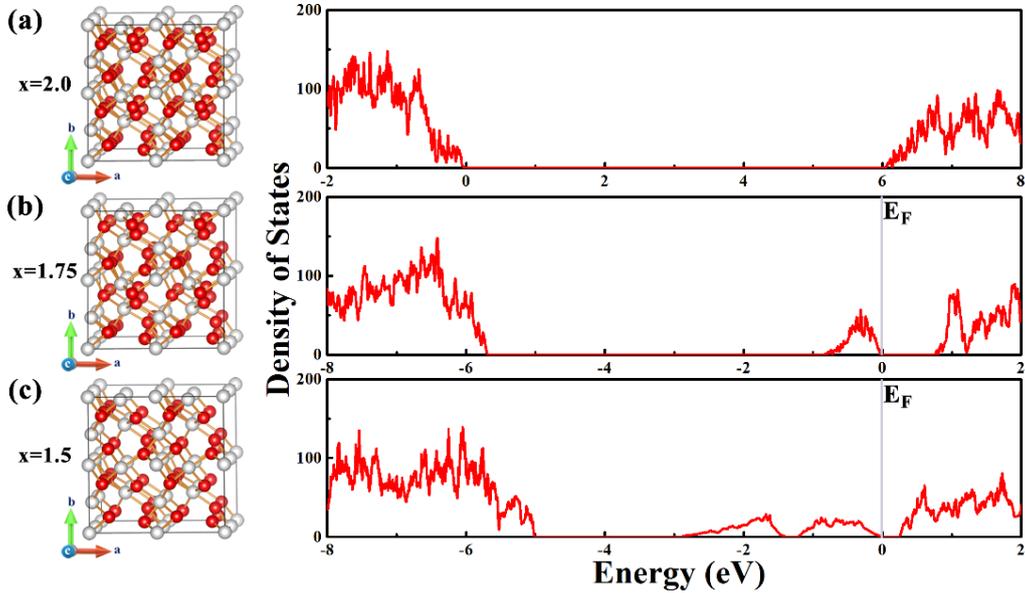

**Figure 10**. The crystal structures and density of states of (a) *m*-HfO$_2$, (b) *m*-HfO$_{1.75}$ and (c) *m*-HfO$_{1.5}$.

Although the conductivity of *m*-HfO$_x$ varies monotonously with O content, the realization of multi-value storage and continuous conductivity modulation of *m*-HfO$_x$ is seriously hindered by the mobile characteristics of Vo. As shown in **Figure 11(a)**, due to the poor symmetry of *m*-HfO$_2$, there are 9 inequivalent Vo migration paths. Distinct from the analysis in **Sect. 3.3.1**, we here only focus on migration between neighbor sites, with any individual hopping distance limited to within 3 Å. This is because RS within *m*-HfO$_x$ does not involve too high driving force, but a phase transition requires drastic structural changes, such that the paths in **Sect. 3.3.1** are all beyond 4 Å. For the sake of conductance modulation, the results in **Figure 11(b)** show that,

Vo in *m*-HfO$_2$ has the lowest migration barrier under Path-8 (1.72 eV). Moreover, continuous migration solely through Path-8 is possible, without invoking other paths. Therefore, CFs can preferably be created along this path in the *c*-axis direction. The Vo migration barriers for other paths lie between 2.26 eV and 3.00 eV. However, when electrons de-trap from a neutral Vo to generate charged Vo$^{2+}$ under the applied electric field, the migration barrier of each path decreases significantly, and the migration barrier of Path-1 is as low as 0.34 eV. The migration barriers of Path-5, Path-6, Path-8 and Path-9 are lower than 1.00 eV. These paths altogether actually cover migration in all directions, thus Vo in *m*-HfO$_2$ is prone to diffusion/drift, which will lead to instability of O content in CFs as well as poor stability and consistency of the HRS. Diffusion can be achieved along multiple paths with mixed directions. Therefore, it is difficult to achieve stable multi-value storage and continuous conductance modulation.

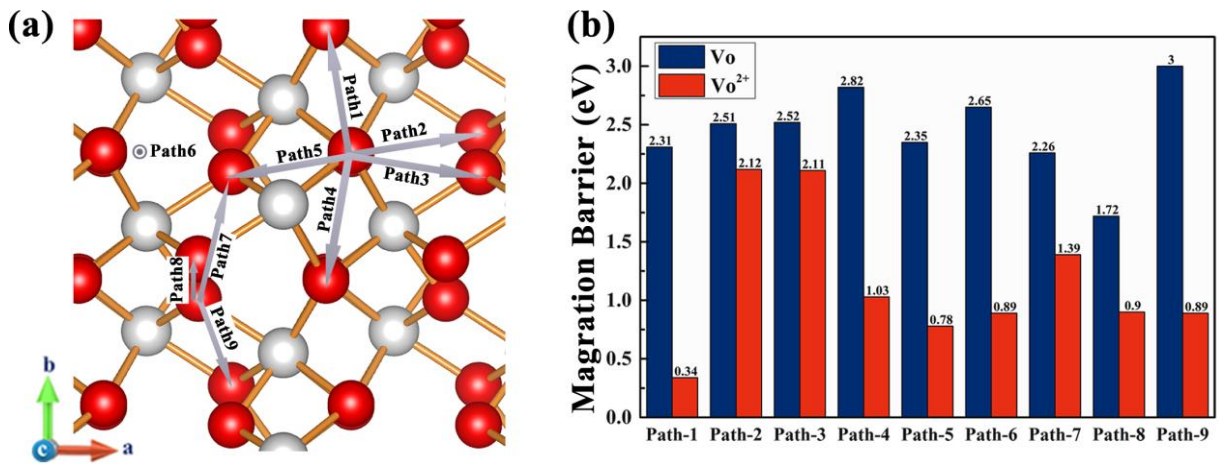

**Figure 11**. (a) The schematic diagram of Vo migration path in *m*-HfO$_2$; (b) the histogram of Vo migration barriers under different paths.

### 3.3.3 RS within *P*4$_2$/*nmc*-HfO$_x$

For 1.0<*x*<1.5, there are phases that are derivatives of γ-HfO$_{1.5}$ and α-HfO$_{1.5}$, respectively. Conductive paths of γ-HfO$_{1.5}$ composition have been observed in experiments [34], but CFs of α-HfO$_{1.5}$ composition have never been discovered in memristors. In addition, derivatives of α-HfO$_{1.5}$ can be regarded as derivatives of γ-HfO$_{1.5}$ with more Vo's (*cf.* **Figure 6**). Therefore, these CF candidates will be globally referred to as γ-HfO$_x$. In the *x* range of interests, the structures and density of states of γ-HfO$_x$ are shown in **Figure 12**. Obviously, the density of states of γ-HfO$_{1.5}$ is similar to that of m-HfO$_{1.5}$, and the defect states introduced by Vo are distributed below CBM. With the decrease of O content, the number of shallow defect states increases. For $x < 1.25$, γ-HfO$_x$ becomes metallic. The conductivity of γ-HfO$_{1.5}$ is similar to that of m-HfO$_{1.5}$, which indicates that the change of CF structure from m-HfO$_{1.5}$ to γ-HfO$_{1.5}$ will not lead to the sudden change in conductivity of HfO$_2$-based memristor. In addition, the phase transition temperature between γ-HfO$_{1.5}$ and m-HfO$_{1.5}$ is estimated to be merely 62 K, and reversible transition may be realized in the actual operation process. Theoretically, when O content in CFs is maintained at $1 < x < 2$, hafnia-based memristor can realize continuous conductivity modulation with large adjustable range (from an insulated state to a metallic state) by adjusting the oxygen content in CFs.

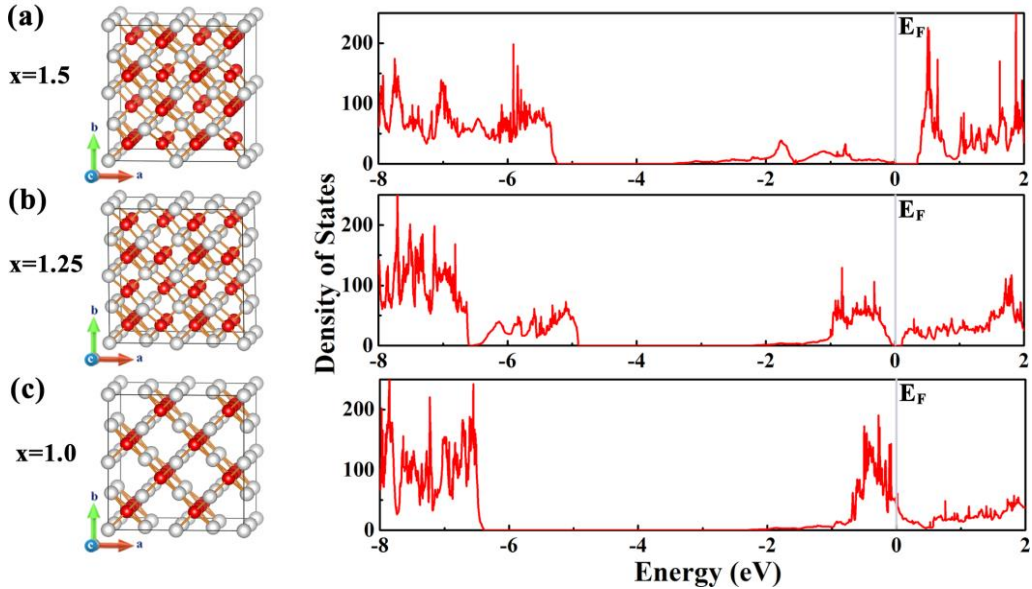

**Figure 12**. The crystal structures and density of states of (a) $P4_2/nmc$-HfO$_{1.5}$, (b) $P4_2/nmc$-HfO$_{1.25}$ and (c) $P4_2/nmc$-HfO.

However, like m-HfO$_x$, the stability of Vo in γ-HfO$_{1.5}$ is also poor. As shown in **Figure 13**, Vo has only two inequivalent paths in γ-HfO$_{1.5}$ due to its higher symmetry. The migration barriers of V$_O$ along Path-1 and Path-2 are 0.83 eV and 1.22 eV, respectively, both lower than that in m-HfO$_2$. Different from m-HfO$_2$, the migration barrier of Vo$^{2+}$ is higher than that of Vo (1.17 eV for Path-1 and 1.82 eV for Path-2), but the migration rate is still suppose to be fast. Therefore, CFs composed of γ-HfO$_x$ still can hardly provide stable low and intermediate resistance states for hafnia-based memristor.

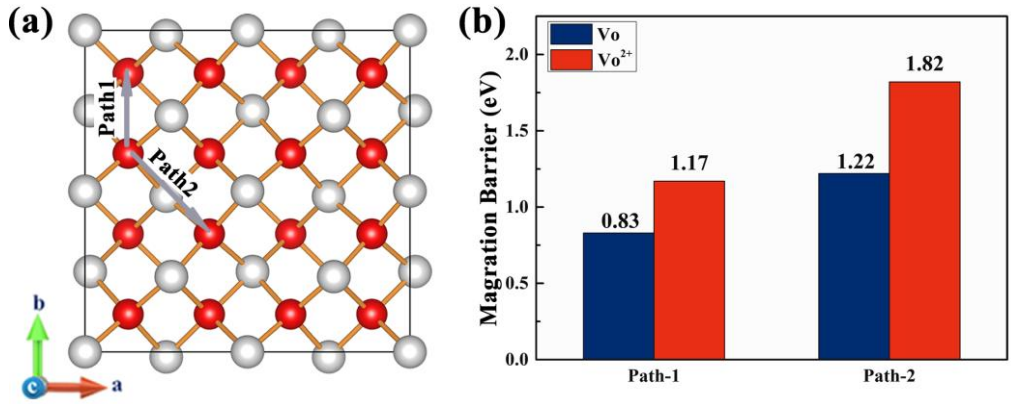

**Figure 13.** (a) The schematic diagram of Vo migration path in $P4_2/nmc$-HfO$_{1.5}$; (b) the histogram of Vo migration barriers under different paths.

### 3.3.4 RS involving $h$-HfO$_x$

The atomic structure and density of states of $h$-HfO$_x$ under various O contents are shown in **Figure 14**, all of which are strong metals and contain many interstitial positions for O. Among the highly reduced HfO$_x$ phases with $x<1.0$, $h$-Hf$_6$O has been proved to be a stable phase at room temperature and pressure [41], whose hexagonal structure is shown in **Figure 14(d)**. CFs with the $h$-HfO$_x$ composition clearly indicate a low resistance state (LRS) of the memristor, and $h$-Hf$_6$O has been widely discovered as the CF composition in the experiment of Zhang *et al*. [34]. Similar to α-HfO$_{1.5}$, $h$-HfO$_x$ requires absorbing a number of O atoms so as to be ruptured. The robust nature of $h$-HfO$_x$ leads to violent SET/RESET processes, and hafnia memristors in this mode exhibit obvious sharp-transition binary characteristics.

Mao *et al.* carried out a comprehensive study on the migration barriers of O atoms in $h$-HfO$_x$ under various O contents, giving barriers between 2.26 eV and 3.64 eV [46], significantly higher than in $m$-HfO$_{1.5}$ and γ-HfO$_{1.5}$. This greatly strengthens the stability of the CF structure. On the other hand, the O content in $h$-HfO$_x$ has little influence on its conductivity characteristics [46], rendering the LRS level robust if constructed from $h$-HfO$_x$. Although multi-value storage and continuous conductance modulation can be realized by controlling the size of the CFs [61], CFs remain highly conductive and the linearly adjustable range of conductance is limited [60].

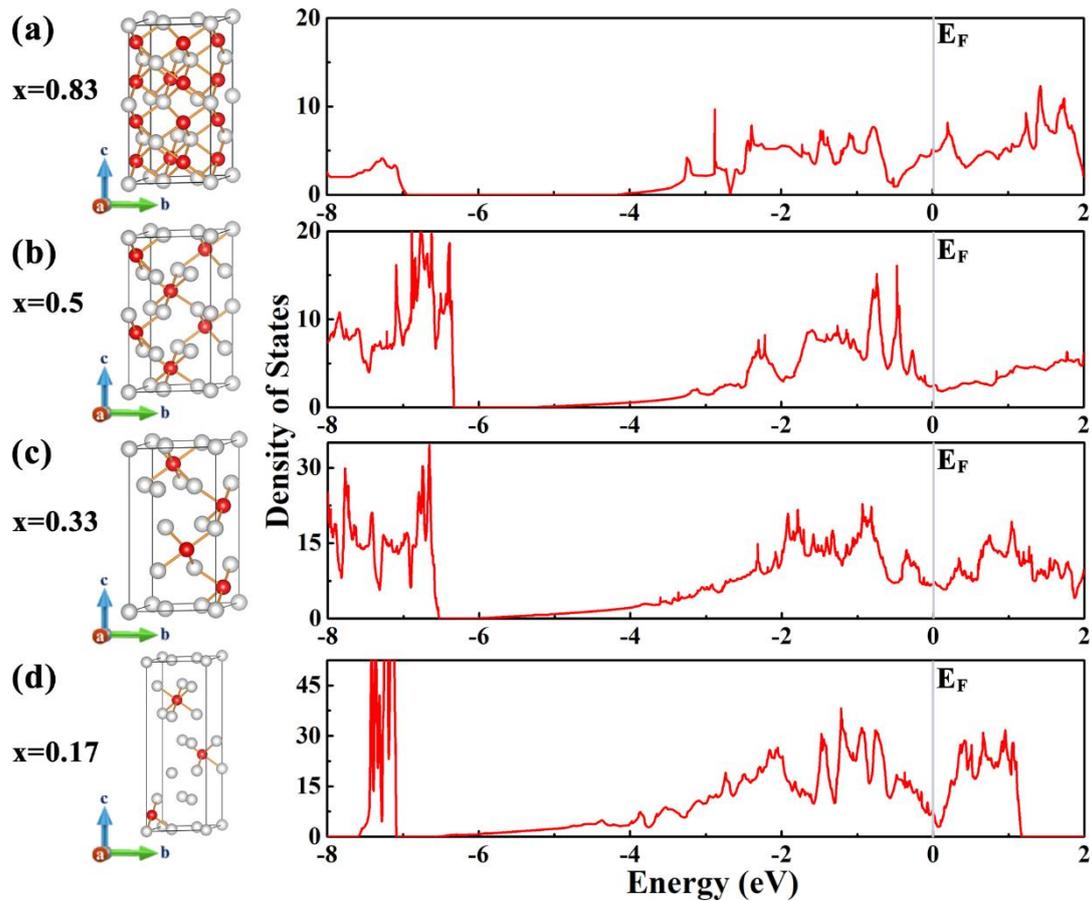

**Figure 14**. The crystal structures and density of states of (a) $h$-$Hf_6O_5$, (b) $h$-$Hf_2O$, (c) $h$-$Hf_2O$, (d) $h$-$Hf_6O$.

### 3.4 The size effect of CFs

The above analysis implies that gradual conductance modulation is possible in hafnia memristors by restricting the CF composition to monoclinic or tetragonal phases, rather than incurring the highly reduced hexagonal phases. The modulation is ideally achieved through controlling the excitation barriers from the defective states. However, experimentally it is difficult to obtain a wide range of continuous conductance modulation in undoped hafnia [24,60,62–66]. In addition to the poor thermal stability of $V_O$'s, the CF size effect in $m$-$HfO_x$ or $\gamma$-$HfO_x$ compositions is another key factor, namely, to what dimension the CFs of $m$-$HfO_x$ or $\gamma$-$HfO_x$ structure can remain stable. According to thermodynamic data [34], a sufficiently large $m$-$HfO_x$ or $\gamma$-$HfO_x$ CF (1.0<$x$<2.0) tend to decompose into $m$-$HfO_2$ and $h$-

Hf$_6$O, but small CFs of such compositions may become stable due to the established interfaces to their chemical environment. Therefore, this section is devoted to the stability of CF models consisting of *m*-HfO$_x$ with various sizes (shown in **Figures 15(a)-15(c)**), as well as CFs composed of *h*-HfO$_x$ with the same V$_O$ concentrations (shown in **Figures 15(d)-15(f)**). All CFs are embedded in a 7×1×7 *m*-HfO$_2$ supercell.

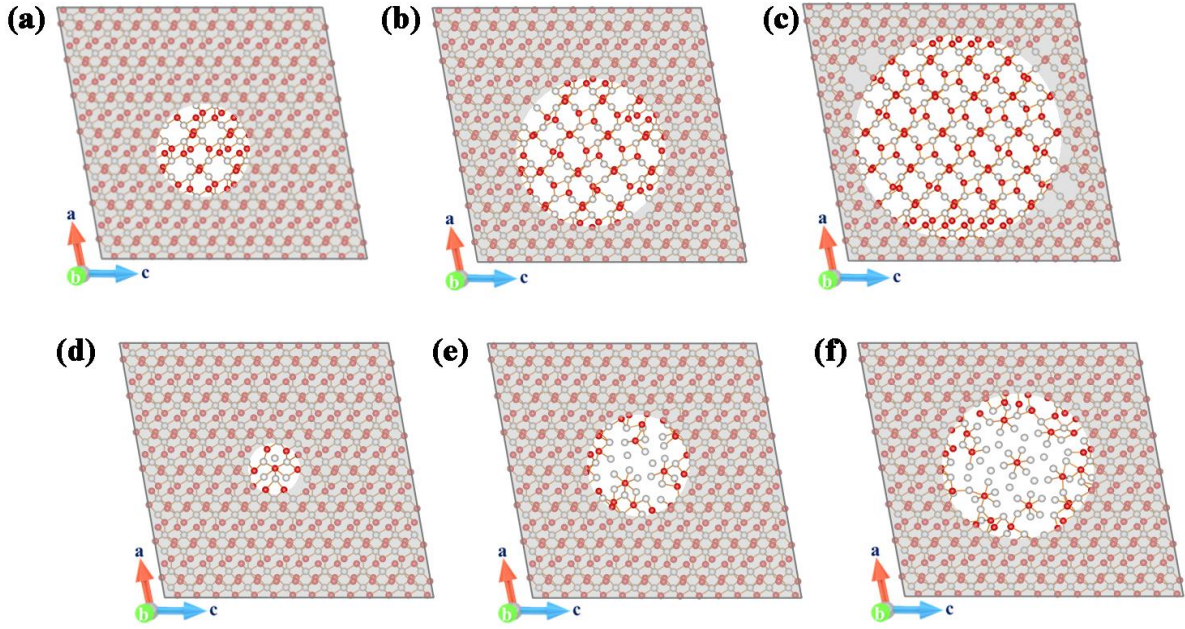

**Figure 15**. The structure diagrams of CF in *m*-HfO$_{1.5}$ with Vo content at (a) 2.8%, (b)11.2% and (c) 25.3%; the structure diagrams of CF in *h*-Hf$_6$O with Vo content at (a) 2.8%, (b)11.2% and (c) 25.3%. Dielectric regions outside the CF are faded to some extent to emphasize the CF location.

In **Figures 15**, only those models with the same V$_O$ content could be compared, *i.e.*, (a) with (d), (b) with (e), and (c) with (f). Since the hexagonal phase can be regarded as the consequence of V$_O$ aggregation, while in *m*-HfO$_x$ the V$_O$'s are still apart from each other, we here define a cohesive energy of the V$_O$ as

$$E_c = \frac{E_{\text{h-Hf}_6O} - E_{\text{m-HfO}_{1.5}}}{N_{V_O}}$$

where $E_{\text{h-Hf}_6O}$ is the formation energy of *h*-Hf$_6$O CFs, $E_{\text{m-HfO}_{1.5}}$ is the formation energy of *m*-HfO$_{1.5}$ CFs, and $N_{V_O}$ is the amount of V$_O$ per supercell. The formation energy of a CF is

$$E_{\text{form}} = E_{\text{model}} + N_{V_O}\mu_O - nE_{\text{HfO}_2}$$

where $E_{model}$ is the energy after model relaxation, $\mu_O$ is the chemical potential of oxygen element, $E_{HfO_2}$ is the energy of ground state $m$-HfO$_2$, and $n$ accounts for the size of the model supercell.

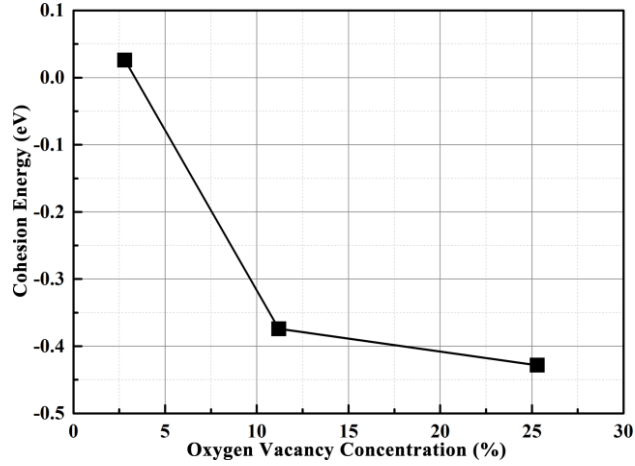

**Figure 16**. The average cohesion energy of CFs in $h$-Hf$_6$O at various V$_O$ contents.

Therefore, $E_c$ reflects the energy required for the CF structure to evolve from $m$-HfO$_{1.5}$ to $h$-Hf$_6$O. When the V$_O$ concentration is merely 2.8%, corresponding to ~1 nm diameter of $m$-HfO$_{1.5}$ CFs, the $E_c$ value is positive (0.03 eV), indicating that $m$-HfO$_{1.5}$ CFs are more stable. When the V$_O$ concentration rises to 11.2% (the diameter of $m$-HfO$_{1.5}$-CFs reaching ~2 nm), $E_c$ decreases rapidly to be a negative value -0.37 eV, predicting a spontaneous transformation of the CFs into $h$-Hf$_6$O. As the V$_O$ concentration further increases, $E_c$ does not change significantly, but still listing $h$-Hf$_6$O as the favorable CF composition. Hence, CFs composed of $m$-HfO$_{1.5}$ would become unstable as its diameter exceeds 1 nm, collapsing to $h$-Hf$_6$O. This explains the CF shell structure identified recently, where $h$-Hf$_6$O CFs were surrounded by $m$-HfO$_2$ (or sometimes t-HfO$_2$) from the microscope observation [34]. The cohesive tendency of V$_O$'s thus prefers the $h$-Hf$_6$O CF composition, as long as the SET/RESET operations are not so constrained, explaining the usually discovered binary RS characteristics in undoped hafnia.

### 3.5 Effect of Mg doping on the RS mechanism of hafnia memristor

The above analysis points out two reasons why satisfactory gradual conductance modulation is difficult to be achieved in undoped hafnia. On the one hand, the low and somehow isotropic migration barriers for $V_O$ add to the instability of the resistance levels of the HRS. On the other hand, the strong accumulation tendency of $V_O$ and $V_O$ chains [40] facilitate the formation of $h$-$Hf_6O$ CFs, promoting the binary switching mode. Element doping is, however, a convenient and effective means to adjust the $V_O$ migration kinetics in hafnia. For instance, very recently we have reported the encouraging role of Mg doping in improving the resistance level stability in hafnia memristors [60]. Here we review the impact of Mg doping in terms of the migration kinetics arguments.

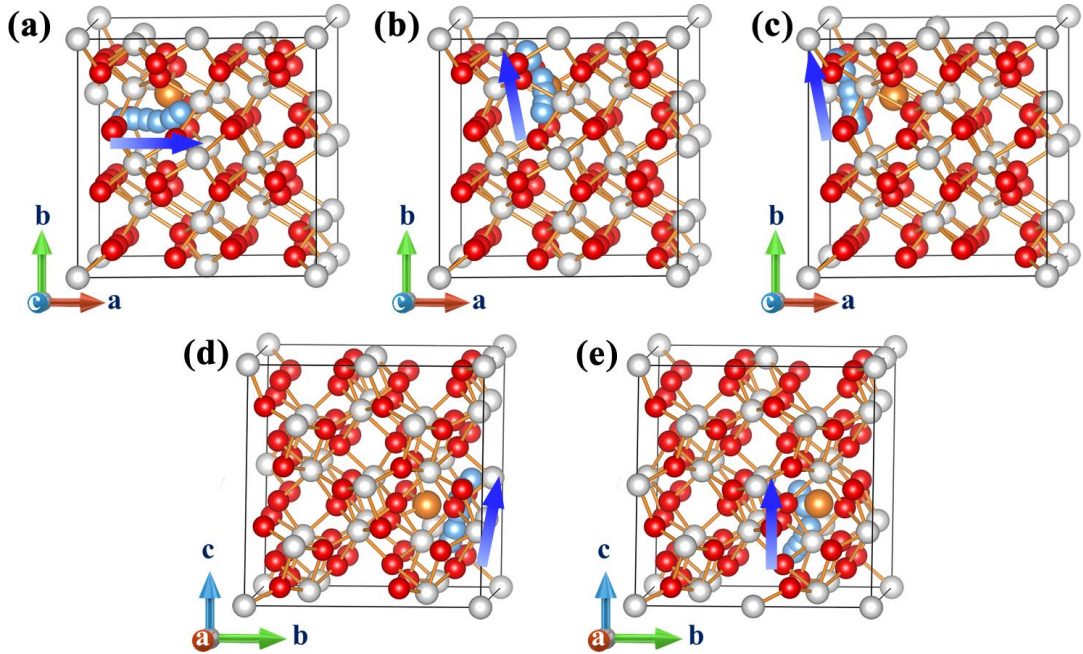

**Figure 17**. Migration paths diagram of Vo near Mg dopant in different directions. The green balls plus the corresponding arrows indicate the paths.

It is known that Mg tends to occupy interstitial locations in hafnia [67], where the favorite location is indicated by the orange atoms in **Figure 17**. Our focus is the $V_O$'s surrounding an Mg dopant. Their migration barriers in several directions, as illustrated in **Figure 17**, are illustrated in **Figure 18**. Compared with undoped hafnia, the migration barriers of $V_O$ and $V_O^{2+}$ near the Mg dopant have both increased dramatically, reaching 2.24 eV ~ 2.69 eV ($V_O$) and 1.82 eV ~ 2.80 eV ($V_O^{2+}$), respectively. This would improve the resistance level stability in hafnia, explaining the robust multi-level RS phenomena observed in Mg:$HfO_x$ memristors [60]. On the other hand, $V_O^{2+}$ surrounding Mg tends to migrate along Path-b2 (1.82 eV barrier), but all other paths present barriers no less than 2.16 eV. Consequently, Mg doping introduces some anisotropy in Vo diffusion/migration, favoring the CFs along the direction of Path-b2. However, then the horizontal migration from or into the CFs becomes more difficult. To a certain extent, such anisotropic migration tends to hinder the merging of small CFs into a strong CF, delaying the possible phase transition into $h$-Hf$_6$O. These two features are helpful for the memristor to work in the gradual conductance modulation mode. Similar theoretical analysis may also be used for other element doping in hafnia-based memristors.

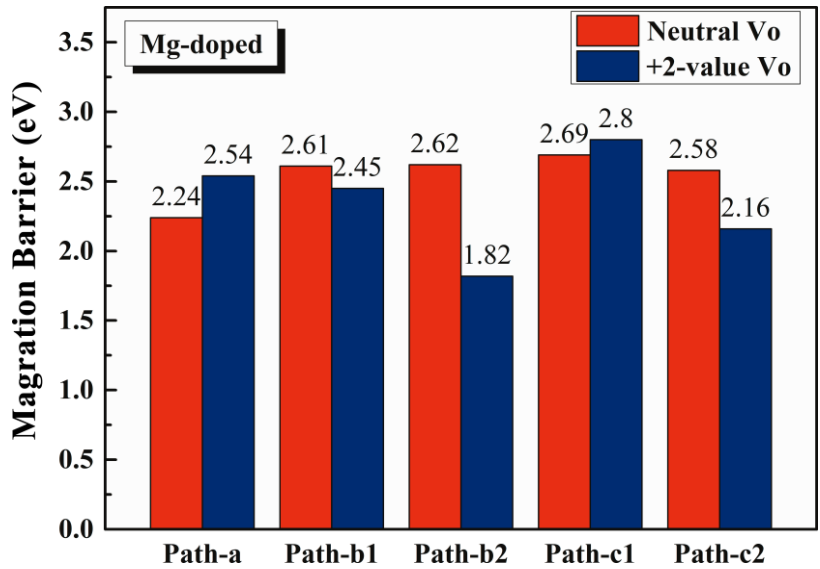

**Figure 18**. Migration barriers of Vo and Vo$^{2+}$ near Mg dopant in different directions.

## 4. Conclusion

We have explored the rule of structural evolution in HfO$_2$-based memristors, under the condition of memristor operation modes. The possibility of continuous and gradual conductance modulation is discussed for the sake of neuromorphic applications. The following conclusions can be drawn.

(1) HfO$_x$ underwent a transition from the monoclinic phase to tetragonal phases, and eventually to hexagonal phases during the reduction process in a memristor. The Hf sub-system remains in an *f.c.c.*-like structure for 1<x<2, and transitions between various sub-oxide phases in this regime can be achieved through O vacancy migration. The conductance modulation is gradual by controlling the density and energy level of defect states. However, the Hf sub-system tends to transform into *h.c.p.*-like structure for x<1, and the rupture and reformation of hexagonal conductive filaments render abrupt binary switching characteristics.

(2) The previously predicted $P\bar{4}m2$-phase $Hf_2O_3$ is found to consist of an *f.c.c.*-like Hf sub-lattice with half O anions located at the rock salt sites but half O anions at the fluorite sites. The transition barrier from monoclinic $HfO_2$ to $P\bar{4}m2$-phase $Hf_2O_3$ is high under the memristor operating conditions, thus a defective $Hf_2O_3$ derived by introducing O vacancies in $P4_2/nmc$-$HfO_2$, and another predicted *Ibam*-$Hf_2O_3$ phase are more likely to emerge in $HfO_x$-memristors.

(3) The recently observed $I4_1/amd$-phase is the most stable structure for the HfO stoichiometry at room temperature, and is similar to a $P4_2/mmc$-HfO phase. Both derive from $P4_2/nmc$-$HfO_2$ through O vacancy introduction, which is possible in memristor operations. The $P\bar{6}2m$ phase HfO can be transformed from or to $I4_1/amd$-HfO by temperature effect, which sets up a bridge between *f.c.c.* Hf-lattice and *h.c.p.* Hf-lattice. Nevertheless, such transition cannot be achieved through a mild O vacancy migration process in a memristor environment. Hence, the resistive switching between $Hf/Hf_6O$ with *h.c.p.* Hf-lattice and $HfO_x$ (x>1) with quasi-*f.c.c.* Hf-lattice inevitably involves an abrupt transition.

(4) The difficulty in obtaining continuous and robust conductance modulation in undoped $HfO_x$ lies in that, the migration of O vacancies in monoclinic or tetragonal $HfO_x$ structures is both fast and nearly isotropic, which facilitates the merging of O vacancies and O vacancy chains into strong filaments. Our calculation shows that strong filaments tend to become hexagonal due to a phase transition at around *x*=1, leading to typical binary switching characteristics. The effect Mg doping is to introduce a certain degree of anisotropy in O vacancy diffusion/migration, which hinders the formation of hexagonal filaments, and thus stabilizes the mode of gradual conductance modulation.

**Acknowledgement**

This work was supported by the National Key R&D Program of China under Grant No. 2019YFB2205100.

**References**

[1] S. Ginnaram, J. T. Qiu, and S. Maikap, Role of the Hf/Si Interfacial Layer on the High Performance of MoS$_2$-Based Conductive Bridge RAM for Artificial Synapse Application, IEEE Electron Device Lett. **41**, 709 (2020).

[2] X. Hong, D. J. Loy, P. A. Dananjaya, F. Tan, C. Ng, and W. Lew, Oxide-Based RRAM Materials for Neuromorphic Computing, J Mater Sci **53**, 8720 (2018).

[3] S. Maikap and W. Banerjee, In Quest of Nonfilamentary Switching: A Synergistic Approach of Dual Nanostructure Engineering to Improve the Variability and Reliability of Resistive Random-Access-Memory Devices, Adv. Electron. Mater. **6**, 2000209 (2020).

[4] B. Song, R. Cao, H. Xu, S. Liu, H. Liu, and Q. Li, A HfO$_2$/SiTe Based Dual-Layer Selector Device with Minor Threshold Voltage Variation, Nanomaterials **9**, 408 (2019).

[5] D. Ielmini and H.-S. P. Wong, In-Memory Computing with Resistive Switching Devices, Nat Electron **1**, 333 (2018).

[6] S. Balatti, S. Ambrogio, and D. Ielmini, Normally-off Logic Based on Resistive Switches—Part I: Logic Gates, IEEE Trans. Electron Devices **62**, 1831 (2015).

[7] Z. Li, B. Tian, K.-H. Xue, B. Wang, M. Xu, H. Lu, H. Sun, and X. Miao, Coexistence of Digital and Analog Resistive Switching With Low Operation Voltage in Oxygen-Gradient HfO$_x$ Memristors, IEEE Electron Device Lett. **40**, 1068 (2019).

[8] Z. Wang, M. Yin, T. Zhang, Y. Cai, Y. Wang, Y. Yang, and R. Huang, Engineering Incremental Resistive Switching in TaO$_x$ Based Memristors for Brain-Inspired Computing, Nanoscale **8**, 14015 (2016).

[9] M. Mao, S. Yu, and C. Chakrabarti, Design and Analysis of Energy-Efficient and Reliable 3-D ReRAM Cross-Point Array System, IEEE Trans. VLSI Syst. **26**, 1290 (2018).

[10] A. Zaffora, D.-Y. Cho, K.-S. Lee, F. Di Quarto, R. Waser, M. Santamaria, and I. Valov, Electrochemical Tantalum Oxide for Resistive Switching Memories, Adv. Mater. **29**, 1703357 (2017).

[11] S. Choi, S. H. Tan, Z. Li, Y. Kim, C. Choi, P.-Y. Chen, H. Yeon, S. Yu, and J. Kim, SiGe Epitaxial Memory for Neuromorphic Computing with Reproducible High Performance Based on Engineered Dislocations, Nature Mater **17**, 335 (2018).

[12] G. W. Burr, R. M. Shelby, A. Sebastian, S. Kim, S. Kim, S. Sidler, K. Virwani, M. Ishii, P. Narayanan, A. Fumarola, L. L. Sanches, I. Boybat, M. Le Gallo, K. Moon, J. Woo, H. Hwang, and Y. Leblebici, Neuromorphic Computing Using Non-Volatile Memory, Advances in Physics: X **2**, 89 (2017).

[13] P.-H. Chen, K.-C. Chang, T.-C. Chang, T.-M. Tsai, C.-H. Pan, T.-J. Chu, M.-C. Chen, H.-C. Huang, I. Lo, J.-C. Zheng, and S. M. Sze, Bulk Oxygen–Ion Storage in Indium–Tin–Oxide Electrode for Improved Performance of $HfO_2$-Based Resistive Random Access Memory, IEEE Electron Device Lett. **37**, 280 (2016).

[14] Y. Li, K.-S. Yin, M.-Y. Zhang, L. Cheng, K. Lu, S.-B. Long, Y. Zhou, Z. Wang, K.-H. Xue, M. Liu, and X.-S. Miao, Correlation Analysis between the Current Fluctuation Characteristics and the Conductive Filament Morphology of $HfO_2$-Based Memristor, Appl. Phys. Lett. **111**, 213505 (2017).

[15] Y. Wang, Q. Liu, S. Long, W. Wang, Q. Wang, M. Zhang, S. Zhang, Y. Li, Q. Zuo, J. Yang, and M. Liu, Investigation of Resistive Switching in Cu-Doped $HfO_2$ Thin Film for Multilevel Non-Volatile Memory Applications, Nanotechnology **21**, 045202 (2010).

[16] T.-M. Tsai, C.-H. Wu, K.-C. Chang, C.-H. Pan, P.-H. Chen, N.-K. Lin, J.-C. Lin, Y.-S. Lin, W.-C. Chen, H. Wu, N. Deng, and H. Qian, Controlling the Degree of Forming Soft-Breakdown and Producing Superior Endurance Performance by Inserting BN-Based Layers in Resistive Random Access Memory, IEEE Electron Device Lett. **38**, 445 (2017).

[17] D. Acharyya, A. Hazra, and P. Bhattacharyya, A Journey towards Reliability Improvement of $TiO_2$ Based Resistive Random Access Memory: A Review, Microelectronics Reliability **54**, 541 (2014).

[18] A. Prakash, J. Park, J. Song, J. Woo, E.-J. Cha, and H. Hwang, Demonstration of Low Power 3-Bit Multilevel Cell Characteristics in a $TaO_x$-Based RRAM by Stack Engineering, IEEE Electron Device Lett. **36**, 32 (2015).

[19] L. Zhao, H.-Y. Chen, S.-C. Wu, Z. Jiang, S. Yu, T.-H. Hou, H.-S. P. Wong, and Y. Nishi, Multi-Level Control of Conductive Nano-Filament Evolution in $HfO_2$ ReRAM by Pulse-Train Operations, Nanoscale **6**, 5698 (2014).

[20] L. Yang, C. Kuegeler, K. Szot, A. Ruediger, and R. Waser, The Influence of Copper Top Electrodes on the Resistive Switching Effect in $TiO_2$ Thin Films Studied by Conductive

Atomic Force Microscopy, Appl. Phys. Lett. **95**, 013109 (2009).

[21] F. Zahoor, T. Z. Azni Zulkifli, and F. A. Khanday, Resistive Random Access Memory (RRAM): An Overview of Materials, Switching Mechanism, Performance, Multilevel Cell (Mlc) Storage, Modeling, and Applications, Nanoscale Res Lett **15**, 90 (2020).

[22] Z.-H. Wu, K.-H. Xue, and X.-S. Miao, Filament-to-Dielectric Band Alignments in $TiO_2$ and $HfO_2$ Resistive RAMs, J Comput Electron **16**, 1057 (2017).

[23] K. Moon, M. Kwak, J. Park, D. Lee, and H. Hwang, Improved Conductance Linearity and Conductance Ratio of 1T2R Synapse Device for Neuromorphic Systems, IEEE Electron Device Lett. **38**, 1023 (2017).

[24] W. He, H. Sun, Y. Zhou, K. Lu, K. Xue, and X. Miao, Customized Binary and Multi-Level $HfO_{2-x}$-Based Memristors Tuned by Oxidation Conditions, Sci Rep **7**, 10070 (2017).

[25] T. Kim, H. Kim, J. Kim, and J.-J. Kim, Input Voltage Mapping Optimized for Resistive Memory-Based Deep Neural Network Hardware, IEEE Electron Device Lett. **38**, 1228 (2017).

[26] P. Huang, X. Y. Liu, B. Chen, H. T. Li, Y. J. Wang, Y. X. Deng, K. L. Wei, L. Zeng, B. Gao, G. Du, X. Zhang, and J. F. Kang, A Physics-Based Compact Model of Metal-Oxide-Based RRAM DC and AC Operations, IEEE Trans. Electron Devices **60**, 4090 (2013).

[27] S. Brivio, J. Frascaroli, and S. Spiga, Role of Metal-Oxide Interfaces in the Multiple Resistance Switching Regimes of $Pt/HfO_2/TiN$ Devices, Appl. Phys. Lett. 6 (2015).

[28] U. Celano, L. Goux, R. Degraeve, A. Fantini, O. Richard, H. Bender, M. Jurczak, and W. Vandervorst, Imaging the Three-Dimensional Conductive Channel in Filamentary-Based Oxide Resistive Switching Memory, Nano Lett. **15**, 7970 (2015).

[29] S. Kumar, Z. Wang, X. Huang, N. Kumari, N. Davila, J. P. Strachan, D. Vine, A. L. D. Kilcoyne, Y. Nishi, and R. S. Williams, Conduction Channel Formation and Dissolution Due to Oxygen Thermophoresis/Diffusion in Hafnium Oxide Memristors, 6 (n.d.).

[30] Y. Yang, X. Zhang, L. Qin, Q. Zeng, X. Qiu, and R. Huang, Probing Nanoscale Oxygen Ion Motion in Memristive Systems, Nat Commun **8**, 15173 (2017).

[31] J. Yin, F. Zeng, Q. Wan, F. Li, Y. Sun, Y. Hu, J. Liu, G. Li, and F. Pan, Adaptive Crystallite Kinetics in Homogenous Bilayer Oxide Memristor for Emulating Diverse Synaptic Plasticity, Adv. Funct. Mater. **28**, 1706927 (2018).

[32] G. Bersuker, D. C. Gilmer, D. Veksler, P. Kirsch, L. Vandelli, A. Padovani, L. Larcher, K. McKenna, A. Shluger, V. Iglesias, M. Porti, and M. Nafría, Metal Oxide Resistive Memory Switching Mechanism Based on Conductive Filament Properties, Journal of Applied Physics **110**, 124518 (2011).

[33] K.-H. Xue, P. Blaise, L. R. C. Fonseca, G. Molas, E. Vianello, B. Traoré, B. De Salvo, G. Ghibaudo, and Y. Nishi, Grain Boundary Composition and Conduction in $HfO_2$: An Ab Initio Study, Appl. Phys. Lett. **102**, 201908 (2013).

[34] Y. Zhang, G.-Q. Mao, X. Zhao, Y. Li, M. Zhang, Z. Wu, W. Wu, H. Sun, Y. Guo, L. Wang, X. Zhang, Q. Liu, H. Lv, K.-H. Xue, G. Xu, X. Miao, S. Long, and M. Liu, Evolution of the Conductive Filament System in $HfO_2$-Based Memristors Observed by Direct Atomic-Scale Imaging, Nat Commun **12**, 7232 (2021).

[35] K. P. McKenna, Optimal Stoichiometry for Nucleation and Growth of Conductive Filaments in $HfO_x$, Modelling Simul. Mater. Sci. Eng. **22**, 025001 (2014).

[36] T. V. Perevalov and D. R. Islamov, Atomic and Electronic Structures of the Native Defects Responsible for the Resistive Effect in $HfO_2$: Ab Initio Simulations, Microelectronic Engineering **216**, 111038 (2019).

[37] Y. Dai, Z. Pan, F. Wang, and X. Li, Oxygen Vacancy Effects in $HfO_2$ -Based Resistive Switching Memory: First Principle Study, AIP Advances **6**, 085209 (2016).

[38] F. Xu, B. Gao, Y. Xi, J. Tang, H. Wu, and H. Qian, Atomic-Device Hybrid Modeling of Relaxation Effect in Analog RRAM for Neuromorphic Computing, in 2020 IEEE International Electron Devices Meeting (IEDM) (IEEE, San Francisco, CA, USA, 2020), p. 13.2.1-13.2.4.

[39] L. Sementa, L. Larcher, G. Barcaro, and M. Montorsi, Ab Initio Modelling of Oxygen Vacancy Arrangement in Highly Defective $HfO_2$ Resistive Layers, Phys. Chem. Chem. Phys. **19**, 11318 (2017).

[40] K.-H. Xue and X.-S. Miao, Oxygen Vacancy Chain and Conductive Filament Formation in Hafnia, J. Appl. Phys. 11 (2018).

[41] J. Zhang, A. R. Oganov, X. Li, K.-H. Xue, Z. Wang, and H. Dong, Pressure-Induced Novel Compounds in the Hf-O System from First-Principles Calculations, Phys. Rev. B **92**, 184104 (2015).

[42] B. Puchala and A. Van der Ven, Thermodynamics of the Zr-O System from First-Principles Calculations, Phys. Rev. B **88**, 094108 (2013).

[43] Kan-Hao Xue, B. Traore, P. Blaise, L. R. C. Fonseca, E. Vianello, G. Molas, B. De Salvo, G. Ghibaudo, B. Magyari-Kope, and Y. Nishi, A Combined Ab Initio and Experimental Study on the Nature of Conductive Filaments in Pt/HfO$_2$/Pt Resistive Random Access Memory, IEEE Trans. Electron Devices 61, 1394 (2014).

[45] K. Z. Rushchanskii, S. Blügel, and M. Ležaić, Routes for Increasing Endurance and Retention in HfO$_2$ -Based Resistive Switching Memories, Phys. Rev. Materials **2**, 115002 (2018).

[46] G.-Q. Mao, K.-H. Xue, Y.-Q. Song, W. Wu, J.-H. Yuan, L.-H. Li, H. Sun, S. Long, and X.-S. Miao, Oxygen Migration around the Filament Region in HfO$_x$ Memristors, AIP Advances **9**, 105007 (2019).

[47] N. Kaiser, T. Vogel, A. Zintler, S. Petzold, A. Arzumanov, E. Piros, R. Eilhardt, L. Molina-Luna, and L. Alff, Defect-Stabilized Substoichiometric Polymorphs of Hafnium Oxide with Semiconducting Properties, ACS Appl. Mater. Interfaces acsami.1c09451 (2021).

[48] X. Luo, W. Zhou, S. V. Ushakov, A. Navrotsky, and A. A. Demkov, Monoclinic to Tetragonal Transformations in Hafnia and Zirconia: A Combined Calorimetric and Density Functional Study, Phys. Rev. B **80**, 134119 (2009).

[49] P. Hohenberg and W. Kohn, Inhomogeneous Electron Gas, Phys. Rev. **136**, B864 (1964).

[50] W. Kohn and L. J. Sham, Self-Consistent Equations Including Exchange and Correlation Effects, Phys. Rev. **140**, A1133 (1965).

[51] G. Kresse and J. Furthmüller, Efficiency of Ab-Initio Total Energy Calculations for Metals and Semiconductors Using a Plane-Wave Basis Set, Computational Materials Science **6**, 15 (1996).

[52] G. Kresse and J. Furthmüller, Efficient Iterative Schemes for Ab Initio Total-Energy Calculations Using a Plane-Wave Basis Set, Phys. Rev. B **54**, 11169 (1996).

[53] L. G. Ferreira, M. Marques, and L. K. Teles, Approximation to Density Functional Theory for the Calculation of Band Gaps of Semiconductors, Phys. Rev. B **78**, 125116 (2008).

[54] K.-H. Xue, J.-H. Yuan, L. R. C. Fonseca, and X.-S. Miao, Improved LDA-1/2 Method for Band Structure Calculations in Covalent Semiconductors, Computational Materials

Science **153**, 493 (2018).

[55] G. Henkelman, B. P. Uberuaga, and H. Jónsson, A Climbing Image Nudged Elastic Band Method for Finding Saddle Points and Minimum Energy Paths, The Journal of Chemical Physics **113**, 9901 (2000).

[56] L. Zhu, J. Zhou, Z. Guo, and Z. Sun, Metal–Metal Bonding Stabilized Ground State Structure of Early Transition Metal Monoxide TM–MO (TM = Ti, Hf, V, Ta), J. Phys. Chem. C **120**, 10009 (2016).

[57] J.-H. Yuan, G.-Q. Mao, K.-H. Xue, N. Bai, C. Wang, Y. Cheng, H. Lyu, H. Sun, X. Wang, and X. Miao, Ferroelectricity in $HfO_2$ from a Chemical Perspective, ArXiv:2201.00210 [Cond-Mat] (2022).

[58] E. Hildebrandt, J. Kurian, M. M. Müller, T. Schroeder, H.-J. Kleebe, and L. Alff, Controlled Oxygen Vacancy Induced p-Type Conductivity in $HfO_{2-x}$ Thin Films, Appl. Phys. Lett. **99**, 112902 (2011).

[59] R. R. Manory, T. Mori, I. Shimizu, S. Miyake, and G. Kimmel, Growth and Structure Control of $HfO_{2-x}$ Films with Cubic and Tetragonal Structures Obtained by Ion Beam Assisted Deposition, Journal of Vacuum Science & Technology. A, Vacuum, Surfaces, and Films **20**, 549 (2002).

[60] L.-H. Li, K.-H. Xue, L.-Q. Zou, J.-H. Yuan, H. Sun, and X. Miao, Multilevel Switching in Mg-Doped $HfO_x$ Memristor through the Mutual-Ion Effect, Appl. Phys. Lett. **119**, 153505 (2021).

[61] S.-M. Park, H.-G. Hwang, J.-U. Woo, W.-H. Lee, S.-J. Chae, and S. Nahm, Improvement of Conductance Modulation Linearity in a $Cu^{2+}$-Doped $KNbO_3$ Memristor through the Increase of the Number of Oxygen Vacancies, ACS Appl. Mater. Interfaces **12**, 1069 (2020).

[62] S. Chandrasekaran, Improving Linearity by Introducing Al in $HfO_2$ as a Memristor Synapse Device, 10 (2019).

[63] B. Gao, L. Liu, and J. Kang, Investigation of the Synaptic Device Based on the Resistive Switching Behavior in Hafnium Oxide, Progress in Natural Science: Materials International **25**, 47 (2015).

[64] S. Kim, Engineering Synaptic Characteristics of $TaO_x/HfO_2$ Bi-Layered Resistive Switching Device, 9 (2018).

[65] S. Kim, J. Chen, Y.-C. Chen, M.-H. Kim, H. Kim, M.-W. Kwon, S. Hwang, M. Ismail, Y. Li, X.-S. Miao, Y.-F. Chang, and B.-G. Park, Neuronal Dynamics in $HfO_x/AlO_y$-Based Homeothermic Synaptic Memristors with Low-Power and Homogeneous Resistive Switching, Nanoscale **11**, 237 (2019).

[66] S. Roy, G. Niu, Q. Wang, Y. Wang, Y. Zhang, H. Wu, S. Zhai, P. Shi, S. Song, Z. Song, Z.-G. Ye, C. Wenger, T. Schroeder, Y.-H. Xie, X. Meng, W. Luo, and W. Ren, Toward a Reliable Synaptic Simulation Using Al-Doped $HfO_2$ RRAM, ACS Appl. Mater. Interfaces **12**, 10648 (2020).

[67] D. Duncan, B. Magyari-Köpe, and Y. Nishi, Properties of Dopants in $HfO_x$ for Improving the Performance of Nonvolatile Memory, Phys. Rev. Applied **7**, 034020 (2017).

# Supplementary Material for

# Hafnia for analog memristor: Influence of stoichiometry and crystalline structure

Li-Heng Li,[1,2] Kan-Hao Xue,[1,2*] Jun-Hui Yuan,[1,2] Ge-Qi Mao,[1,2] and Xiangshui Miao[1,2]

[1]School of Integrated Circuits, School of Optical and Electronic Information, Huazhong University of Science and Technology, Wuhan 430074, China

[2]Hubei Yangtze Memory Laboratories, Wuhan 430205, China

*Corresponding Author, E-mail: xkh@hust.edu.cn (K.-H. Xue)

## Supplementary Note 1.

The calculation formula of the binding energy is as follows:

$$E_{binding} = \frac{E_{HfO_x} - N_{Hf}\mu_{Hf} - \frac{N_O}{2}\mu_{O_2}}{N_{Hf}}$$

Where, $E_{HfO_x}$ is the calculated total energy of the HfO$_x$ structure, $N_{Hf}$ represents the number of Hf atoms in the structure, $\mu_{Hf}$ is the chemical potential of Hf atoms, $N_O$ is the number of O atoms in the structure, $\mu_{O_2}$ is the chemical potential of O$_2$.

Table 1 The binding energies of the different $m$-HfO$_{1.5}$ structures

|  | Case-4 | Case-5 | Case-7 | Case-9 |
|---|---|---|---|---|
| Binding Energy (eV/f.u.) | -7.70 | -7.63 | -7.58 | -7.65 |

## Supplementary Note 2: Vibration entropy calculation

To obtain the vibration entropy, we adopted the quantum harmonic oscillator model,[1] where in each degree of the freedom one has

$$\varepsilon_n = \left(n + \frac{1}{2}\right)h\nu \qquad (S1)$$

Given that the vibration frequencies have been obtained through density functional theory together with the finite difference method, the zero point energy is simply

$$ZPE = \frac{h}{2}\sum_{i=1}^{3N} v_i \quad (S2)$$

where the sum should be carried out over all degrees of freedom. For each degree of the freedom, the partition function becomes

$$Z^{(i)} = \sum_{n=0}^{\infty} exp\left(-\frac{\varepsilon_n^{(i)}}{kT}\right) = \frac{exp(-\frac{x^{(i)}}{2})}{1 - exp(-x^{(i)})} \quad (S3)$$

where

$$x^{(i)} \equiv \frac{hv_i}{kT} \quad (S4)$$

Hence, the vibration entropy for each degree of freedom should be

$$S_{vib}^{(i)} = kInZ^{(i)} + kT\frac{\partial InZ^{(i)}}{\partial T} = \frac{kx^{(i)}}{exp(x^{(i)} - 1)} - kIn[1 - exp(-x^{(i)})] \quad (S5)$$

and the total vibration entropy is

$$S_{vib} = \sum_{i=1}^{3N} S_{vib}^{(i)} \quad (S6)$$

## Supplementary Note 3.

Table 2 The bond lengths of Hf-O and Bader charges of Hf atom.

| Structure | The bond length of Hf-O (Å) | Bader charge of Hf atom |
|---|---|---|
| $I4_1/amd$ | 2.10/2.22 | +1.36e/+1.50e |
| $P4_2/mmc$ | 2.14 | +1.49e |
| $P\bar{6}2m$ | 2.13/2.25 | +1.39e/+1.44e |
| $Fm\bar{3}m$ | 2.29 | +1.63e |

**Supplementary Note 4.**

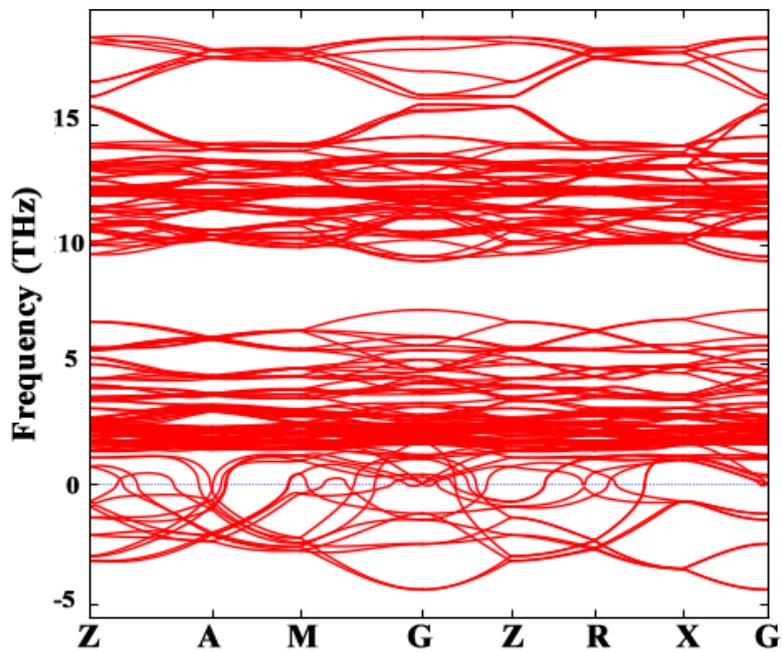

**Figure S1** The phonon spectrum of γ-HfO$_{0.75}$

**References**

1　L. Lv, Z. Li, K.-H. Xue, Y. Ruan, X. Ao, H. Wan, X. Miao, B. Zhang, J. Jiang, C. Wang and K. (Ken) Ostrikov, *Nano Energy*, 2018, **47**, 275–284.